\documentclass{IEEEtran}

\usepackage{amsmath}
\usepackage{mathtools}
\usepackage{graphicx}
\usepackage{enumerate}
\usepackage{url}
\usepackage{subfigure}
\usepackage{algorithm}
\usepackage{algpseudocode}
\usepackage{stfloats}
\usepackage{lineno}
\usepackage{hyperref}
\usepackage{bm}
\usepackage{bbm}
\usepackage{amssymb}
\usepackage{enumerate}
\usepackage{xcolor}
\usepackage{float}
\usepackage{cite}
\usepackage{tabularx}
\usepackage{tabu}
\usepackage{multicol}
\usepackage{multirow}
\usepackage{colortbl,booktabs,threeparttable}
\usepackage{dcolumn}
\usepackage{flushend}
\usepackage{soul}
\usepackage{ragged2e}

\graphicspath{{Figures/}}

\newcommand{\rscl}[1]{\mathrm{#1}}  
\renewcommand{\vec}[1]{\bm #1}
\newcommand{\rvec}[1]{\mathbf{#1}}
\newcommand{\mat}[1]{\bm #1}

\newcommand{\vech}[1]{\hat{\bm #1}}

\renewcommand{\math}[1]{\hat{\bm #1}}

\newcommand{\matb}[1]{\bar{\bm #1}}

\newcommand{\bb}[1]{\mathbb{#1}}
\renewcommand{\cal}[1]{\mathcal{#1}}

\newcommand{\Tr}{\operatorname{Tr}}

\newcommand{\R}{\mathbb{R}}
\newcommand{\C}{\mathbb{C}}
\newcommand{\E}{\mathbb{E}}
\newcommand{\D}{\mathbb{D}}
\renewcommand{\P}{\mathbb{P}}
\newcommand{\Q}{\mathbb{Q}}
\newcommand{\Ph}{\hat{\mathbb{P}}}

\newcommand{\Pb}{\bar{\mathbb{P}}}

\renewcommand{\d}{\mathrm{d}}
\renewcommand{\th}{\text{th}}

\newcommand{\T}{\mathsf{T}}
\renewcommand{\H}{\mathsf{H}}
\renewcommand{\st}{\text{s.t.}}

\DeclareMathOperator*{\argmax}{argmax}
\DeclareMathOperator*{\argmin}{argmin}

\newcommand{\defeq}{\coloneqq}
\newcommand{\stp}{\hfill $\square$}     

\newcommand{\quotemark}[1]{``#1”}

\definecolor{hl-bg-color}{RGB}{255,255,215}
\sethlcolor{hl-bg-color} 
\definecolor{new-magenta}{RGB}{255,0,255}
\soulregister\cite7
\soulregister\citep7
\soulregister\citet7
\soulregister\ref7
\soulregister\eqref7
\newcommand*{\HIGHLIGHT}{}
\ifdefined\HIGHLIGHT

\else

\fi

\newcommand*{\IEEE}{}
\ifdefined\IEEE

    \newtheorem{highlight}{{Highlight}}
    \newtheorem{definition}{{Definition}}

\fi

\begin{document}
\newpage
\title{Robust Processing and Learning: Principles, Methods, and Wireless Applications}

\author{Shixiong Wang,~
        Wei Dai,
        Li-Chun Wang,~\IEEEmembership{Fellow,~IEEE},
        and Geoffrey Ye Li,~\IEEEmembership{Fellow,~IEEE} 
\thanks{S. Wang, W. Dai, and G. Y. Li are with the Department of Electrical and Electronic Engineering, Imperial College London, London SW7 2AZ, United Kingdom (E-mail: s.wang@u.nus.edu; wei.dai1@imperial.ac.uk; geoffrey.li@imperial.ac.uk). L. Wang is with the Department of Electrical and Computer Engineering, National Yang Ming Chiao Tung University, Hsinchu 30010, Taiwan (E-mail: wang@nycu.edu.tw).
}
}

\maketitle

\begin{abstract}
This tutorial-style overview article examines the fundamental principles and methods of robustness, using wireless sensing and communication (WSC) as the narrative and exemplifying framework. WSC forms the backbone of modern information acquisition and exchange, e.g., radar grids and mobile networks, respectively. However, real-world deployments of such systems face ubiquitous uncertainties arising from imperfect models, time-varying environments, limited data, adversarial perturbations, and hardware impairments. If ignored, WSC systems may undergo severe performance degradation and resource inefficiency. Hence, robust design principles are indispensable to ensure reliable and efficient operations under these uncertainties. In this article, we provide a tutorial-style overview of robust methods for WSC from a signal processing perspective, where signal processing, in a modern sense, refers widely to data analytics and decision making for informatics by leveraging statistical, optimization, and machine learning approaches. First, we formalize the conceptual and mathematical foundations of robustness, highlighting the interpretations and relations across robust statistics, optimization, and machine learning. Key techniques, such as robust estimation and testing, distributionally robust optimization, and regularized and adversary training, are investigated. Together, the costs of robustness in system design, for example, the compromised nominal performances and the extra computational burdens, are discussed. Second, we review recent robust signal processing solutions for WSC that address model mismatch, data scarcity, adversarial perturbation, and distributional shift. Specific applications include robust ranging-based localization, modality sensing, channel estimation, receive combining, waveform design, and federated learning. Through this effort, we aim to introduce the classical developments and recent advances in robustness theory to the general signal processing community, exemplifying how robust statistical, optimization, and machine learning approaches can address the uncertainties inherent in WSC systems.
\end{abstract}


\section{Introduction}

The 20th century witnessed the advent of radio technologies that fundamentally transformed humans' ways of environmental perception and information exchange. Guglielmo Marconi’s pioneering transatlantic radio transmissions in 1901 marked the birth of wireless communications, while Robert Watson-Watt’s demonstration in 1935 revealed the capability of radio waves for remote object detection and ranging. Since the 1930s, signal processing techniques enabled by statistics and optimization, such as optimal detection, estimation, filtering, and resource allocation, have played an instrumental role in improving operational performance and approaching the information-theoretic limits of WSC systems. This traditional model-driven (i.e., physics-informed) paradigm persisted until the 2010s, when advanced WSC systems began embracing data-driven machine learning methods to tackle complex signal processing tasks in dynamic, uncertain, and large-scale environments, for example, processing massive data and making trustworthy decisions. Here, trustworthiness broadly includes diverse merits of solutions, such as high performance, computational efficiency, interpretability (e.g., transparency and accountability), reliability (e.g., adaptivity and robustness), and sustainability (e.g., power efficiency). This tutorial-style overview is concerned with reliable signal processing for WSC systems under uncertain conditions, where signal processing is nowadays broadly interpreted to encompass statistical, optimization, and machine learning approaches for informatics, in either a physics-informed or data-driven manner.

\begin{figure}[!htbp]
    \centering
    \includegraphics[width=6.5cm]{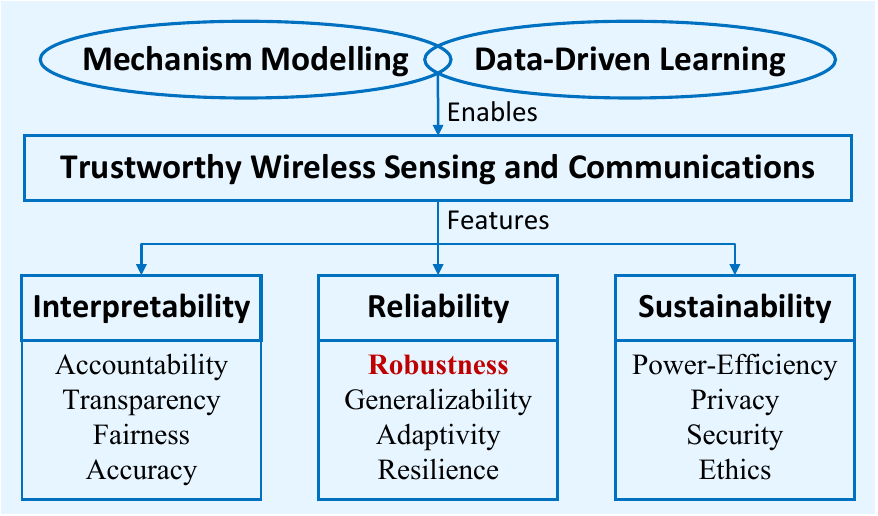}
    \caption{Among the various aspects of trustworthy wireless sensing and communications, robustness holds a critical position.}
    \label{fig:ITP}
\end{figure}

Uncertainties arise when our knowledge departs from the underlying truth, and they are ubiquitous across nearly all scientific procedures, such as modeling, computational, and experimental processes; see \cite{smith2024uncertainty,kassam1985robust,wang2025uncertainty,gawlikowski2023survey}. In engineering research and practice, it is widely recognized that ``all models are wrong, although some are useful", as noted by George Box in 1976 \cite{box1976science}. This is because models are humans' epistemic simplifications of physical reality, which always omit some aspects of the underlying true system, process, situation, or event. George Box's phrase can be arguably generalized: all numerical computations are inaccurate, although some are credible; all experiments are imperfect, although some are informative. This generalization is rooted in the following two facts: numerical computations involve approximations, discretization, and truncation errors, so they are never perfectly accurate; experiments are subject to aleatory measurement noises, sensor calibration errors, and environmental variability. In WSC system engineering and operation, specific instances of uncertainties include channel modeling errors due to imperfect knowledge of radio propagation, statistical inference insufficiency due to data scarcity and limited computations, distributional shifts due to environmental variations, adversarial perturbations due to interference and attacks, among many others \cite{wang2025uncertainty}. Uncertainties can degrade the performance of almost all algorithmic procedures in signal processing, leading statistical, optimization, and machine learning solutions to generate unreliable conclusions or decisions. To this end, modern signal processing, viewed from the broader perspectives of statistics, optimization, and machine learning, focuses largely on the reliability against diverse types of uncertainties arising from modeling, computational, and experimental aspects. Technically, reliability against uncertainties can be achieved from two facets: adaptivity and robustness \cite{kassam1985robust,fauss2021minimax,wang2025uncertainty}. The former seeks to accommodate uncertainties by adjusting the decision or design of a system or algorithm in response to varying environments or conditions, so that the system or algorithm can dynamically maintain its (nearly) optimal running performance. The latter, in contrast, focuses on tolerating uncertainties using a predetermined decision or design, such that the system or algorithm can sustain its actual performance within a satisfactory range even in the presence of uncertainties. The essence of adaptivity lies in leveraging updated observations and knowledge to reduce uncertainties and subsequently adjust the system or algorithm's behavior, whereas robustness relies on a fixed decision or design to tolerate uncertainties without requiring any further architectural, computational, or operational adjustments. In practice, when new observations and knowledge are not available, the robustness strategy is indispensable, which constitutes the technical focus of this article; see Fig. \ref{fig:ITP}.

Robustness principles widely apply to statistics, optimization, and machine learning for data analytics and decision making in informatics, and the corresponding technical treatments need to be specifically developed. In statistics, robust estimation, testing, and regression, formally starting from the 1960s by Peter Huber, aim to handle inaccurate modeling assumptions and distributional shifts, such as structural and parametric model mismatches, data perturbations and outliers, to name a few \cite{huber2009robust,loh2024theoretical}. Complications arise if the dimensions of the involved variables are large, which is common in modern statistical inference, for example, channel estimation and receive beamforming for massive multiple-input multiple-output (MIMO) WSC systems \cite{wang2025distributionallycombining}. Computationally efficient algorithms for high-dimensional robust statistics have only been available very recently, since around 2016 \cite{diakonikolas2019robust,liu2023robust,loh2024theoretical}. In operations research, stochastic programming, robust optimization, distributionally robust optimization, and chance-constrained optimization, originating in the 1950s and flourishing in the 1990s, serve as unified modeling frameworks to combat uncertainties in the parameters of decision models \cite{ben2009robust,kuhn2025distributionally}. Technically, robustness is achieved by seeking the worst-case optimality of the objective and the worst-case feasibility of the constraint(s), so that the decision may remain satisfactory for all realizations of uncertain parameters within a predefined uncertainty set. These robustness solutions provide great opportunities for WSC system engineering and operation when the underlying problem can be formulated into an optimization model, for example, beamforming, waveform design, and resource allocation; see, e.g., \cite{liu2024survey,wang2024robust}. In machine learning, facing different kinds of uncertainty issues such as data scarcity and adversarial perturbations, robust training techniques, including regularization, data augmentation, dropout, pruning, gradient clipping, adversarial training, and model aggregation, are proposed to improve generalizability \cite{xu2012robustness,gawlikowski2023survey,braiek2025machine}. Classical robust machine learning evolved closely alongside robust statistics and robust optimization, whereas modern robust machine learning has been characterized by adversarial training since the 2010s. These robust learning approaches lay the foundation for not only traditional regression and classification methods but also modern large-scale generative pre-trained transformer (GPT) models. In WSC systems, typical application examples include robust data-driven channel estimation, robust federated learning under malicious users and noisy communications, and reliable training of wireless foundation models; see, e.g., \cite{balevi2020high,pillutla2022robust,ang2020robust,buffelli2025towards}. 

Despite the broad scope of wireless sensing and communication, whether the two functions are considered separately or jointly \cite{liu2022integrated}, only a few tutorial and overview articles address the principles and methods of robustness against uncertainties. The first introductory yet comprehensive review of uncertainty awareness in WSC system design is presented in \cite{wang2025uncertainty}, whereas recognizable representatives in general robust signal processing include \cite{kassam1985robust}, \cite{zoubir2012robust}, and \cite{fauss2021minimax}. In the mentioned tutorial and overview literature \cite{wang2025uncertainty,kassam1985robust,zoubir2012robust,fauss2021minimax}, robustness is either conceptually implied through specific use cases \cite{kassam1985robust,wang2025uncertainty} or partially treated in narrow technical scopes, i.e., robust estimation under data outliers \cite{zoubir2012robust} and robust testing under distributional shifts \cite{fauss2021minimax}. Although other subtopics relevant to robust WSC exist, e.g., robust adaptive beamforming \cite{vorobyov2013principles}, adversarial machine learning for wireless communications \cite{adesina2022adversarial}, and adversarially robust sensing for autonomous vehicles \cite{modas2020toward}, they still remain context-specific. Therefore, a tutorial-style overview that provides a mathematical formalism of robustness in signal processing, from the broader perspective of robust statistics \cite{diakonikolas2019robust,loh2024theoretical,huber2009robust}, optimization \cite{ben2009robust,kuhn2025distributionally}, and machine learning \cite{gawlikowski2023survey,braiek2025machine}, is developed. In addition, recent application examples illustrating robustness principles and methods in WSC are presented, including robust ranging-based localization \cite{wang2021denoising}, robust modality sensing for autonomous vehicles \cite{modas2020toward}, robust channel estimation using general adversarial networks \cite{balevi2020high}, distributionally robust receive combining \cite{wang2025distributionallycombining}, robust waveform design for integrated sensing and communication \cite{wang2024robust}, and robust federated learning under adversarial users and noisy communications \cite{pillutla2022robust,ang2020robust}. 

\section{Theory of Robustness}
In this section, we present a unifying and coherent narrative of robustness theory. A mathematical formalism of robustness is first introduced, followed by its application to robust statistics, robust optimization, and robust machine learning. WSC-specific motivations and details are exemplified, yet minimized to emphasize the theory’s broad applicability in general engineering.

\subsection{Uncertainty Issues}
Uncertainty refers to a lack of complete certainty or knowledge about a situation, event, process, or system; that is, it indicates the discrepancy between our knowledge and the underlying reality. In the practice of WSC signal processing, uncertainties can widely occur in models, computations, and experiments. Typical representatives include model mismatches, numerical errors, and data inaccuracies, respectively. Such uncertainties can be caused by, for example, incomplete knowledge of physics, limited word length of digital computers, hardware imperfections (e.g., in manufacturing and calibration), and environmental shifts.

Examples of model mismatches include the misspecification of model types (e.g., using a linear model to represent a nonlinear system or a Gaussian distribution to summarize a non-Gaussian dataset), the misidentification of model parameters (e.g., inexact noise statistics, signal powers, or initial conditions), and the approximation of exact models (e.g., using a reduced model for simplicity). Examples of numerical errors include software bugs, hardware failures, rounding errors, quantization errors, and output ambiguities resulting from the random or strategic initialization and termination of iterative algorithms. Examples of data inaccuracies include measurement biases (e.g., using a spatial, temporal, or racial subgroup of data to represent the population), measurement imprecision (e.g., random gross errors and limited sensor resolutions), impulse noises (e.g., large interference or attacks), data scarcity (e.g., limited sampling of the population although unbiased), data aging (e.g., distributional shift due to environmental variations), adversarial perturbations (arising from, e.g., malicious attacks or unconscious interference), etc.

Uncertainties can be categorized as either epistemic or aleatoric. The former arises from a lack of knowledge or incomplete information, which can potentially be reduced by developing better models, collecting more data, or refining measurements (e.g., data bias). The latter, however, reflects inherent randomness or variability that cannot be mitigated through human efforts but can be characterized probabilistically (e.g., thermal noise). The boundary between the two categories is not strict. For example, whether an uncertainty is epistemic or aleatoric can be relative, because obtaining higher-fidelity models or information may be feasible for some researchers but infeasible for others.

In WSC systems, specific examples of uncertainty issues are as follows \cite{wang2025uncertainty}.
\begin{itemize}
    \item \textit{Channel Uncertainties}: Channel uncertainties, which cause misunderstandings of the propagation behavior of wireless signals, arise from mismatched modeling assumptions (e.g., multi-path fading versus fading-free) and insufficient statistical inference (e.g., limited pilots and computational resources). In addition, unpredictable environmental changes, such as moving scatterers and evolving surroundings, further complicate the acquisition of accurate and real-time channel state information.

    \item \textit{Noises, Interferences, and Attacks}: Wireless systems are often exposed to various noise and interference sources. Thermal noises, environmental fluctuations, co-channel and adjacent-channel interference, as well as jamming and impulse noises, can degrade signal quality and system performance. In addition, adversarial data perturbations in machine learning tasks can mislead system behaviors.

    \item \textit{Hardware Imperfections}: Physical devices in wireless systems are never ideal. Aging components, manufacturing variations, nonlinear amplifiers, sensor calibration errors, clock drift, etc., can lead to deviations from expected hardware performance.

    \item \textit{Installation and Configuration Errors}: These errors result from mismatches between the actual and planned installation locations, or between the implemented and prescribed configuration parameters, respectively. Such issues frequently arise during the deployment of sensors and base stations in WSC networks.

    \item \textit{Variability of Network Topology}: Node mobility and availability, networking protocols, node localization errors, and synchronization errors can all cause frequent changes in connectivity and network topology. If these changes are not accurately known to system and algorithm designers, the operating performance of the system and algorithm may be unsatisfactory.

    \item \textit{Variability of Available Resources}: The actual availability of resources, such as power, bandwidth, and computation, often fluctuates. These variations may stem from sharing policies, competing users, or unpredictable system states. Consequently, unreliable resource allocation decisions can be made under incomplete knowledge.

    \item \textit{Data Inaccuracies}: Data-driven components in wireless systems, such as machine learning algorithms, are vulnerable to imperfections of datasets. Issues like biased data, scarce or outdated samples, and incorrect labeling can degrade data quality. Adversarial samples and pilot contamination further compromise data fidelity.
\end{itemize}

All the aforementioned types of uncertainties can cause statistical, optimization, and machine learning approaches for WSC to produce unreliable conclusions or decisions. Therefore, reliable statistical, optimization, and machine learning techniques that account for uncertainties need to be developed.

\subsection{Uncertainty Awareness}
\begin{figure}[!htbp]
	\centering
	 	\includegraphics[width=5cm]{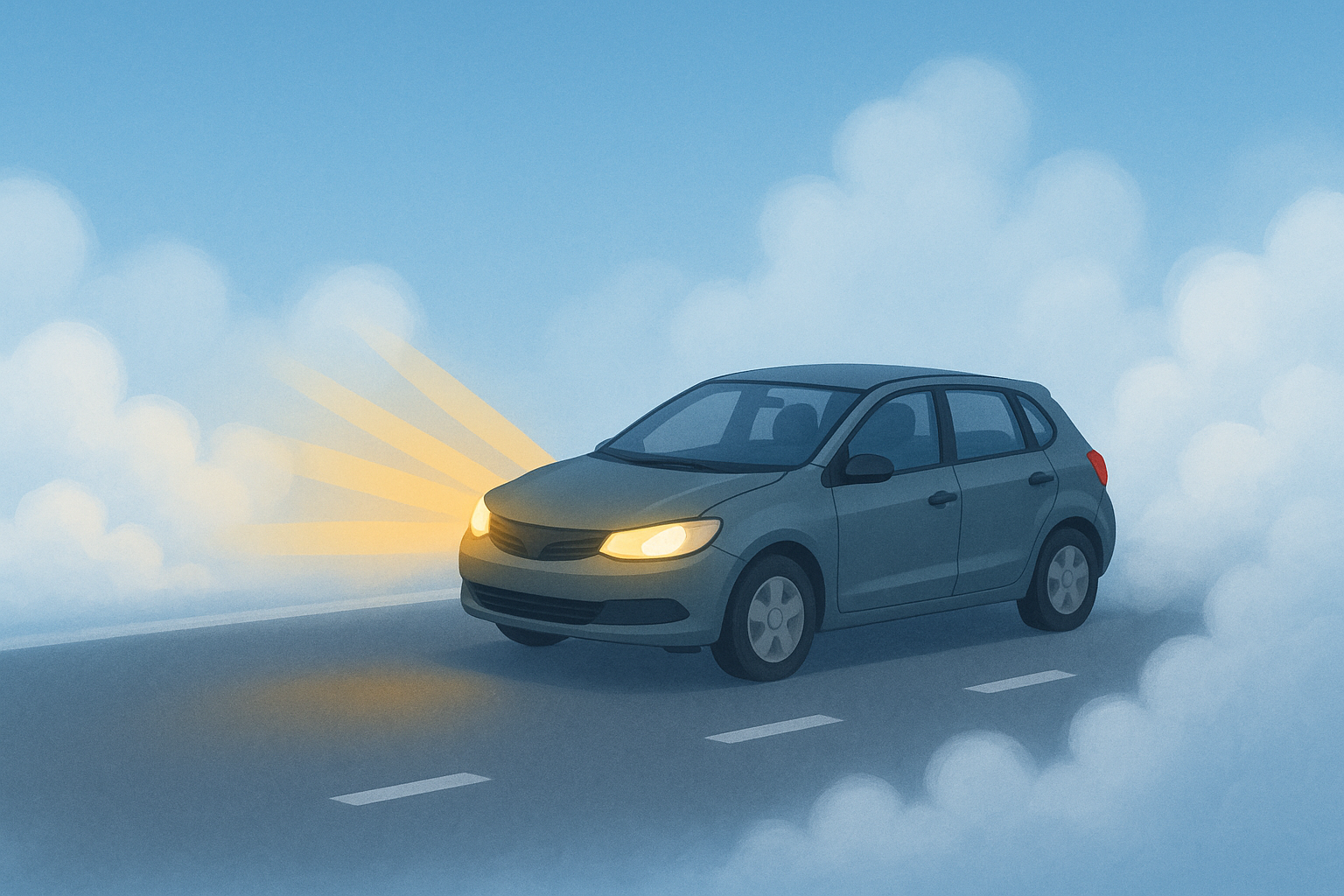}
    
    
	\caption{Driving in uncertain environments is dangerous, so is developing WSC systems without uncertainty awareness. When driving in foggy conditions, an adaptive strategy reduces the uncertainty by sensing the visibility (i.e., fog density) and adjusting the brightness of the fog lights accordingly. In contrast, a robust strategy tolerates the uncertainty by maintaining a consistently low moving speed, regardless of the fog density. (Figure credit: ChatGPT 5.)}
	\label{fig:uncertainty-awareness}
\end{figure}

Facing diverse types of uncertainties, quantifying these uncertainties, drawing uncertainty-informed conclusions (e.g., summarizing a dataset), and making uncertainty-conscious decisions (e.g., improving system performance) are of natural importance. Uncertainty awareness can be broadly defined as the science of identifying, quantifying, propagating, and reducing uncertainties associated with models, numerical algorithms, experiments, and predicted quantities of interest. Technically, uncertainty awareness entails trustworthiness under uncertainties through adaptive and robust treatments. 
A conceptual illustration of the uncertainty issue and uncertainty awareness is given in Fig. \ref{fig:uncertainty-awareness}.

Unlike robustness, adaptivity requires real-time knowledge of evolving conditions and environments. Moreover, additional appropriate resources, such as physical devices and computational power, must be employed to mitigate the induced uncertainties. Therefore, robustness principles and designs are indispensable when neither updated knowledge about uncertainties nor extra resources to reduce them are accessible. 

\subsection{Philosophy and Formalism of Robustness}
This subsection formalizes a unifying and coherent theory of robustness, built on concepts and techniques that are separately discussed across multiple research communities, including statistics \cite{huber2009robust,diakonikolas2019robust,liu2023robust,donoho1988automatic}, optimization \cite{kuhn2025distributionally,romisch2003stability,guo2021statistical}, signal processing \cite{kassam1985robust,wang2025distributionallybeamforming},  and machine learning \cite{xu2012robustness,braiek2025machine}.

We consider an object (e.g., system, process, method, algorithm, etc.) whose real-world performance $h(\vec \xi)$ is determined by a factor $\vec \xi$. Here, $\vec \xi$ may represent input conditions, historical data, or environmental descriptions, among many other quantities that impact the performance measure $h(\vec \xi)$. Without loss of generality, this article supposes that the smaller the value of $h(\vec \xi) \ge 0$, the better the performance of the object. That is, $h(\vec \xi)$ defines a cost function. For example, $h(\vec \xi)$ can denote the positioning error of a radar for a target, while $\vec \xi$ may represent the installation location of the radar. Note that the estimated position of the target depends on the location $\vec \xi$ of the radar.

Assume that the actual value of the factor is $\vec \xi_0$, while the nominal (i.e., assumed or estimated) value is $\vech \xi$. This discrepancy between our knowledge $\vech \xi$ and the underlying reality $\vec \xi_0$ reflects the \textit{uncertainty} associated with the factor $\vec \xi$. Let the uncertainty set of $\vec \xi$ be $\Xi$, which contains all possible realizations of $\vec \xi$. Therefore, $\vech \xi, \vec \xi_0 \in \Xi$. Conceptually, robustness of an object means that its real-world performance $h(\vec \xi)$ remains insensitive to the uncertainty of $\vec \xi$. Mathematically, this implies that the performance deviation $|h(\vec \xi) - h(\vec \xi_0)|$ should be small if the factor perturbation $d(\vec \xi, \vec \xi_0)$ is small, where $d(\cdot, \cdot)$ denotes an appropriate similarity measure between two arguments, e.g., a distance on $\Xi$. As a result, if an assumed parameter $\vech \xi$ can closely approximate the true value $\vec \xi_0$, the corresponding performance degradation $|h(\vech \xi) - h(\vec \xi_0)|$ is minimal. On the other hand, robustness entails that the object can maintain its performance at a satisfactory level $t \ge 0$ even when uncertainties occur. Mathematically, this can be written as $h(\vec \xi) \le t$ for all $\vec \xi \in \Xi$.

\begin{figure}[!htbp]
	\centering
	\subfigure[Robust Object]{
	 	\includegraphics[width=4cm]{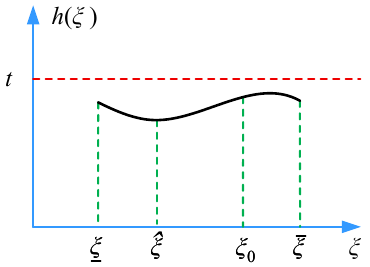}
	}~~~~~~
	\subfigure[Non-robust Object (Jump)]{
	 	\includegraphics[width=4cm]{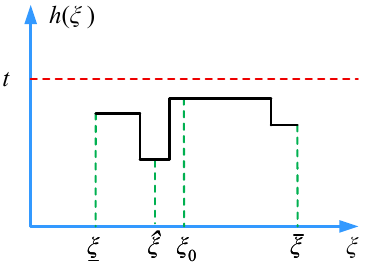}
	}

	\subfigure[Non-robust Object (Excess)]{
	 	\includegraphics[width=4cm]{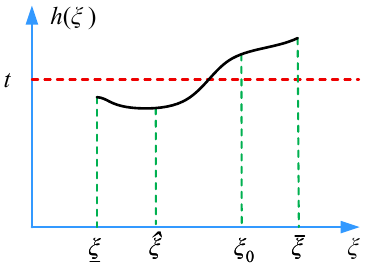}
	}~~~~~~
	\subfigure[Robust Decision]{
	 	\includegraphics[width=4cm]{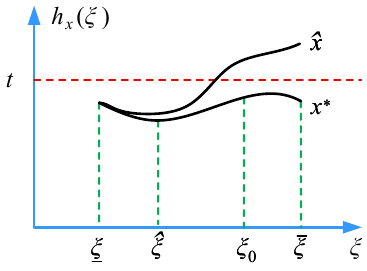}
	}
    
	\caption{Visual illustrations of notions of robustness within the uncertainty set $\Xi \defeq [\underline \xi, \bar \xi]$; note that $\xi_0$ is unknown and can vary on $\Xi$. In (a), the object (e.g., system, method, etc.) is robust against the uncertain factor $\xi$ because the cost does not exceed the tolerance threshold $t$ for all $\xi \in \Xi$. Moreover, a small factor perturbation $|\hat \xi - \xi_0|$ can only cause a small performance deviation. In (b), the object is non-robust because a tiny factor perturbation can lead to a severe performance deviation. In (c), the object is non-robust because the cost can exceed the tolerance threshold for some realizations of $\xi$, e.g., at $\xi_0$. In (d), the decision $x^*$ is robust, while $\hat x$ is non-robust.}
	\label{fig:visualization-robustness}
\end{figure}

In practice, the performance $h_{\vec x}(\vec \xi)$ of an object at the realization $\vec \xi$ is controlled by a decision or design $\vec x \in \cal X$, where $\cal X$ denotes the feasible decision region. Here, $\vec x$ may represent the structure, configuration, operating parameters, etc., of the system, process, or algorithm. For example, $\vec x$ can denote the transmit power of a positioning radar; note that the transmit power of a radar can significantly influence its positioning error. Therefore, the performance $h(\vec x, \vec \xi)$ of the object is jointly determined by both the decision $\vec x$ and the factor $\vec \xi$. In this case, the object is robust at the decision $\vec x$ against the uncertain factor $\vec \xi$ if a small perturbation $d(\vec \xi, \vec \xi_0)$ results in a small deviation $|h_{\vec x}(\vec \xi) - h_{\vec x}(\vec \xi_0)|$ and $h_{\vec x}(\vec \xi) \le t$ for all $\vec \xi \in \Xi_{\vec x}$, where $\Xi_{\vec x}$ may depend on $\vec x$. Accordingly, $\vec x$ is called a \textit{robust decision} against $\vec \xi \in \Xi_{\vec x}$. The subscript of $\Xi_{\vec x}$ should be dropped if $\Xi$ does not rely on $\vec x$; this article assumes this independence for ease of presentation. Visual illustrations of the above notions of robustness are given in Fig. \ref{fig:visualization-robustness}.

On the basis of the above motivations, the formal definitions of robustness are given as follows.

\begin{definition}[Robust Object]
An object (e.g., system, process, method, algorithm, etc.) is $(\epsilon, l, t)$-robust with respect to the cost function $h(\vec \xi)$ and the uncertain factor $\vec \xi \in \Xi$ if\footnote{To simplify the presentation, throughout this tutorial article, we just focus on the most popular ``linear robustness" \cite[Fig.~1]{donoho1988automatic}, \cite{liu2023robust}, \cite{chen2018robust}. The general nonlinear robustness, that is, $|h(\vec \xi) - h(\vec \xi_0)| \le l \cdot e(\epsilon)$ for some nonlinear error functions $e(\epsilon)$, is left to extensive readings \cite{donoho1988automatic,diakonikolas2019robust,lai2016agnostic}. Yet, it requires that $e(\epsilon) \to 0$ as $\epsilon \to 0$.}
\begin{equation}\label{eq:object-robustness}
     |h(\vec \xi) - h(\vec \xi_0)| \le l \cdot \epsilon,~~~\forall \vec \xi: d(\vec \xi, \vec \xi_0) \le \epsilon
\end{equation}
and $h(\vec \xi) \le t$ for all $\vec \xi \in \Xi$. \stp
\end{definition}

In the above definition, the first condition means that the performance deviation of the object is upper-bounded by $l \epsilon$, for all possible assumed parameters $\vech \xi$ that have perturbation levels no larger than $\epsilon$. Given $\epsilon$, the smaller the value of $l$, the more robust the object. Therefore, $l$ can serve as a \textit{quantitative robustness measure} of the object on $\Xi$. In addition, the following mathematical nature of robustness is immediate.

\begin{highlight}[Principle of Robustness]
The robustness of an object essentially favors that the cost function $h(\vec \xi)$ is $l$-Lipschitz continuous and $t$-upper bounded in $\vec \xi$ on $\Xi$.
\stp
\end{highlight}

The Lipschitz continuity, instead of the continuity at the point $\vec \xi_0$, is required because $\vec \xi_0$ can be arbitrary on $\Xi$. Hence, quantitative robustness is an overall property of the cost function $h(\vec \xi)$ on $\Xi$. Similarly, we can define the robustness of a decision $\vec x^*$.

\begin{definition}[Purely Robust Decision]
The decision $\vec x^*$ is purely $(\epsilon, l, t)$-robust with respect to the cost function $h_{\vec x}(\vec \xi)$ and the uncertain factor $\vec \xi \in \Xi$ if
\begin{equation}\label{eq:decision-robustness-pure}
 |h_{\vec x^*}(\vec \xi) - h_{\vec x^*}(\vec \xi_0)| \le l \cdot \epsilon,~~~\forall \vec \xi: d(\vec \xi, \vec \xi_0) \le \epsilon
\end{equation}
and $h_{\vec x^*}(\vec \xi) \le t$ for all $\vec \xi \in \Xi$. \stp
\end{definition}

As the definition suggests, the smaller the value of $l$, the more robust the solution $\vec x^*$ at the perturbation level $\epsilon$. The pure robustness property, however, does not guarantee any performance optimality. At the true value $\vec \xi_0$, the true optimal solution is $\vec x_0 \in \min_{\vec x \in \cal X} h(\vec x, \vec \xi_0)$, so the true optimal cost is $h_{\vec x_0}(\vec \xi_0)$. Consequently, the robust cost $h_{\vec x^*}(\vec \xi_0)$ may be significantly larger than $h_{\vec x_0}(\vec \xi_0)$, compromising the robust decision's performance. As such, in what follows, we adopt the notion of \textit{robust optimality} or \textit{optimal robustness}, but still briefly refer to it as \textit{robustness}.

\begin{definition}[Robust Decision]
The decision $\vec x^*$ is $(\epsilon, l, t)$-robust with respect to the cost function $h_{\vec x}(\vec \xi)$ and the uncertain factor $\vec \xi \in \Xi$ if
\begin{equation}\label{eq:decision-robustness}
 |h_{\vec x^*}(\vec \xi) - h_{\vec x_0}(\vec \xi_0)| \le l \cdot \epsilon,~~~\forall \vec \xi: d(\vec \xi, \vec \xi_0) \le \epsilon
\end{equation}
and $h_{\vec x^*}(\vec \xi) \le t$ for all $\vec \xi \in \Xi$. \stp
\end{definition}

The above definition of robustness mathematically accounts for both robustness and optimality, and thus, is practically favored. When the uncertainty set $\Xi$ is constructed by a ball $\cal B_\epsilon(\vec \xi_0)$ with center $\vec \xi_0$ and radius $\epsilon$, i.e.,
\begin{equation}\label{eq:B0}
\cal B_\epsilon(\vec \xi_0) \defeq \{\vec \xi:d(\vec \xi, \vec \xi_0) \le \epsilon\}, 
\end{equation}
the definitions of robustness can be simplified. We exemplify the robustness definition of a decision as follows. The robustness of an object can be similarly stated.

\begin{definition}[Local Robustness of Decision]
The decision $\vec x^*$ is locally $(\epsilon, l)$-robust with respect to the cost function $h_{\vec x}(\vec \xi)$ and the uncertain factor $\vec \xi \in \cal B_\epsilon(\vec \xi_0)$ if
\begin{equation}\label{eq:local-robustness-measure}
|h_{\vec x^*}(\vec \xi) - h_{\vec x_0}(\vec \xi_0)| \le l \cdot \epsilon,~~~\forall \vec \xi \in \cal B_\epsilon(\vec \xi_0),
\end{equation}
and the prescribed performance threshold $t$ satisfies
\begin{equation}\label{eq:local-satisfaction}
h_{\vec x_0}(\vec \xi_0) + l \cdot \epsilon \le t.
\end{equation}
Here, $l$ is called the \textit{local robustness measure} of $\vec x^*$ with respect to $h_{\vec x}(\vec \xi)$ and $\cal B_\epsilon(\vec \xi_0)$.
\stp
\end{definition}

This definition of the robustness of a decision is local because it demands that the nominal value $\vech \xi$ lies in $\cal B_\epsilon(\vec \xi_0)$, for some $\epsilon \ge 0$. In practice, when the value of $\epsilon$ is difficult to determine or we only know that $\vech \xi$ lies on a general uncertainty set $\Xi$, the above definition becomes less informative. Hence, an alternative definition of robustness, called global robustness, needs to be studied.

\begin{definition}[Global Robustness of Decision]
The decision $\vec x^*$ is globally $(\epsilon, l)$-robust with respect to the cost function $h_{\vec x}(\vec \xi)$ and the uncertain factor $\vec \xi \in \Xi$, if
\begin{equation}\label{eq:global-robustness-measure}
|h_{\vec x^*}(\vec \xi) - h_{\vec x_0}(\vec \xi_0)| \le l \cdot d(\vec \xi, \vec \xi_0),~~~\forall \vec \xi \in \Xi,
\end{equation}
and the prescribed performance threshold $t$ satisfies
\begin{equation}\label{eq:global-satisfaction}
h_{\vec x_0}(\vec \xi_0) + l \cdot d(\vec \xi, \vec \xi_0) \le t,~~~\forall \vec \xi \in \Xi.
\end{equation}
Here, $l$ is called the \textit{global robustness measure} of $\vec x^*$ with respect to $h_{\vec x}(\vec \xi)$ and $d(\vec \xi, \vec \xi_0)$.
\stp
\end{definition}

The global robustness of a decision entails that the performance deviation is upper-bounded by the level of factor perturbation, up to a constant $l$. Similarly, the smaller the value of $l$, the more robust the decision $\vec x^*$. When the level of factor perturbation is known to be no larger than $\epsilon$, the global robustness degenerates to the local robustness because $l \cdot d(\vec \xi, \vec \xi_0)$ is upper-bounded by $l \cdot \epsilon$.

In some engineering applications, the factor $\vec \xi$ randomly takes values on its domain $\Xi$ and is distributed according to the probability distribution $\P_{0, \vec \xi}$. In practice, however, the true distribution $\P_{0, \vec \xi}$ may be unknown, so we assume a nominal (e.g., estimated) distribution $\Ph_{\vec \xi}$ as a surrogate of $\P_{0,\vec \xi}$. This discrepancy between the assumed knowledge $\Ph_{\vec \xi}$ and the underlying reality $\P_{0,\vec \xi}$ leads to a \textit{distributional uncertainty} associated with the random factor $\vec \xi$; that is, a \textit{distributional shift} occurs. For example, in wireless communications, the receive beamforming problem can be formulated as \cite{wang2025distributionallycombining}
\begin{equation}\label{eq:receive-combining}
    \min_{\mat W} \Tr \E_{\P_{0, \rvec x, \rvec s}} [\rvec s - \vec W \rvec x][\rvec s - \vec W \rvec x]^\H,
\end{equation}
where $\Tr$ denotes the matrix trace, $\E$ the expectation operator, $\H$ the conjugate transpose, $\rvec s \in \C^{n_t}$ the transmitted signal, $\rvec x \in \C^{n_r}$ the received signal, and $\vec W \in \C^{n_t \times n_r}$ the receive beamformer; $n_t$ is the number of transmit antennas, while $n_r$ the number of receive antennas. Here, notice that we treat $\vec \xi \defeq [\rvec x; \rvec s]$. In real-world operation, the true joint distribution $\P_{0, \rvec x, \rvec s}$ is unknown, and only the empirical distribution $\Ph_{n, \rvec x, \rvec s}$, constructed using $n$ pilots (i.e., channel training data), is accessible. Therefore, the employment of an approximated distribution $\Ph_{n, \rvec x, \rvec s}$ introduces the distributional uncertainty, or distributional shift, compared to the underlying truth $\P_{0, \rvec x, \rvec s}$. Accordingly, the concepts of distributional robustness need to be investigated, where the performance measure $h(\P_{\vec \xi})$ of an object, or $h_{\vec x}(\P_{\vec \xi})$ of a decision $\vec x$, is a functional of the distribution $\P_{\vec \xi}$ of $\vec \xi$. Let the uncertainty set of $\P_{\vec \xi}$ be $\cal M$; i.e., $\Ph_{\vec \xi}, \P_{0, \vec \xi} \in \cal M$. With a slight abuse of notation, we reuse $d(\P_{\vec \xi}, \P_{0, \vec \xi})$ to denote a similarity measure between two distributions $\P_{\vec \xi}$ and $\P_{0, \vec \xi}$, e.g., a distance on $\cal M$.

\begin{definition}[Distributionally Robust Object]
An object (e.g., system, process, method, algorithm, etc.) is distributionally $(\epsilon, l, t)$-robust with respect to the cost functional $h(\P_{\vec \xi})$ and the uncertain distribution $\P_{\vec \xi} \in \cal M$ if 
\begin{equation}\label{eq:object-dist-robustness}
 |h(\P_{\vec \xi}) - h(\P_{0, \vec \xi})| \le l \cdot \epsilon,~~~\forall \P_{\vec \xi}: d(\P_{\vec \xi}, \P_{0, \vec \xi}) \le \epsilon.
\end{equation}
and $h(\P_{\vec \xi}) \le t$ for all $\P_{\vec \xi} \in \cal M$. \stp
\end{definition}

As indicated, the mathematical nature of distributional robustness can be stated as follows.

\begin{highlight}[Principle of Distributional Robustness]
The distributional robustness of an object essentially favors that the cost functional $h(\P_{\vec \xi})$ is $l$-Lipschitz continuous and $t$-upper bounded in $\P_{\vec \xi}$ on $\cal M$.
\stp
\end{highlight}

When the uncertainty set $\cal M$ is constructed by a ball with center $\P_{0, \vec \xi}$ and radius $\epsilon$, i.e.,
\begin{equation}\label{eq:BP0}
\cal B_\epsilon(\P_{0, \vec \xi}) \defeq \{\P_{\vec \xi}:d(\P_{\vec \xi}, \P_{0, \vec \xi}) \le \epsilon\}, 
\end{equation}
the local distributional robustness of a decision $\vec x^*$ can be defined. In the context of distributional robustness, with a slight abuse of notation, we let $\vec x_0 \in \min_{\vec x \in \cal X} h({\vec x}, \P_{0, \vec \xi})$.

\begin{definition}[Local Distributional Robustness of Decision]
The decision $\vec x^*$ is locally distributionally $(\epsilon, l)$-robust with respect to the cost functional $h_{\vec x}(\P_{\vec \xi})$ and the uncertain distribution $\P_{\vec \xi} \in \cal B_\epsilon(\P_{0, \vec \xi})$, if
\begin{equation}\label{eq:local-dist-robustness-measure}
|h_{\vec x^*}(\P_{\vec \xi}) - h_{\vec x_0}(\P_{0, \vec \xi})| \le l \cdot \epsilon,~~~\forall \P_{\vec \xi} \in \cal B_\epsilon(\P_{0, \vec \xi}),
\end{equation}
and the prescribed performance threshold $t$ satisfies
\begin{equation}\label{eq:local-dist-satisfaction}
h_{\vec x_0}(\P_{0, \vec \xi}) + l \cdot \epsilon \le t.
\end{equation}
Here, $l$ is called the \textit{local distributional robustness measure} of $\vec x^*$ with respect to $h_{\vec x}(\P_{\vec \xi})$ and $\cal B_\epsilon(\P_{0, \vec \xi})$.
\stp
\end{definition}

Similarly, the global distributional robustness of a decision $\vec x^*$ can be defined as follows.

\begin{definition}[Global Distributional Robustness of Decision]
The decision $\vec x^*$ is globally distributionally $(\epsilon, l)$-robust with respect to the cost functional $h_{\vec x}(\P_{\vec \xi})$ and the uncertain distribution $\P_{\vec \xi} \in \cal M$, if
\begin{equation}\label{eq:global-dist-robustness-measure}
|h_{\vec x^*}(\P_{\vec \xi}) - h_{\vec x_0}(\P_{0, \vec \xi})| \le l \cdot d(\P_{\vec \xi}, \P_{0, \vec \xi}),~~~\forall \P_{\vec \xi} \in \cal M,
\end{equation}
and the prescribed performance threshold $t$ satisfies
\begin{equation}\label{eq:global-dist-satisfaction}
h_{\vec x_0}(\P_{0, \vec \xi}) + l \cdot d(\P_{\vec \xi}, \P_{0, \vec \xi}) \le t,~~~\forall \P_{\vec \xi} \in \cal M.
\end{equation}
Here, $l$ is called the \textit{global distributional robustness measure} of $\vec x^*$ with respect to $h_{\vec x}(\P_{\vec \xi})$ and $d(\P_{\vec \xi}, \P_{0, \vec \xi})$.
\stp
\end{definition}

In all previously formalized concepts of robustness, \textit{two-sided robustness} has been focused on. In some statistical problems, the two-sided, or more generally all-directional, robustness is natural. To be specific, let $T: \cal M \to \R^m$ be a statistical functional. In estimating an unknown quantity $\vec \theta_0 \defeq T(\P_{0,\vec \xi})$ (e.g., mean, covariance, regression coefficients) of a clean data distribution $\P_{0, \vec \xi}$, an estimate $\vech \theta \defeq T(\P_{\vec \xi})$ based on the data-generating distribution $\P_{\vec \xi}$ is $(\epsilon,l)$-robust if
\begin{equation}\label{eq:robust-statistics-spirit}
    \|T(\P_{\vec \xi}) - T(\P_{0,\vec \xi})\| \le l \epsilon,~~~\forall \P_{\vec \xi}: d(\P_{\vec \xi}, \P_{0,\vec \xi}) \le \epsilon,
\end{equation}
where $\|\cdot\|$ defines a proper norm on $\R^m$. Note that although we aim to estimate the feature of $\P_{0, \vec \xi}$, the samples are drawn from $\P_{\vec \xi}$, which is contaminated from $\P_{0, \vec \xi}$ under $d(\P_{\vec \xi}, \P_{0,\vec \xi}) \le \epsilon$. Eq. \eqref{eq:robust-statistics-spirit} indicates the robustness of the statistical object $T$, illustrating the spirit of \textit{robust statistics}. However, in some performance-sensitive applications, \textit{one-sided robustness} may be preferred, because for cost functions $h(\vec{\xi})$, $h(\P_{\vec{\xi}})$, $h_{\vec{x}}(\vec{\xi})$, and $h_{\vec{x}}(\P_{\vec{\xi}})$, smaller values indicate better performance. One-sided robustness definitions can be straightforwardly extended from their two-sided counterparts; for example, in robust decision making, one may just replace \eqref{eq:local-robustness-measure} with 
\begin{equation}\label{eq:decision-robustness-one-sided}
 h_{\vec x^*}(\vec \xi) - h_{\vec x_0}(\vec \xi_0) \le l \cdot \epsilon,~~~\forall \vec \xi \in \cal B_\epsilon(\vec \xi_0),
\end{equation}
and \eqref{eq:local-dist-robustness-measure} with
\begin{equation}\label{eq:local-dist-robustness-measure-one-sided}
h_{\vec x^*}(\P_{\vec \xi}) - h_{\vec x_0}(\P_{0, \vec \xi}) \le l \cdot \epsilon,~~~\forall \P_{\vec \xi} \in \cal B_\epsilon(\P_{0, \vec \xi}).
\end{equation}

Decision-theoretically, for both one-sided and two-sided robustness, the technical aim is to minimize the robustness measures $l$. To be specific, two-sided robust decision-making can be formulated as 
\begin{equation}\label{eq:decision-making-robustness-two-sided}
\begin{array}{cll}
 \displaystyle \min_{\vec x \in \cal X, l \ge 0} & l \\
 \st & |h(\vec x, \vec \xi) - h(\vec x_0, \vec \xi_0)| \le l \cdot \epsilon, &\forall \vec \xi \in \cal B_\epsilon(\vec \xi_0), \\
 & h(\vec x_0, \vec \xi_0) + l \cdot \epsilon \le t,
\end{array}
\end{equation}
while two-sided distributionally robust decision-making can be written as
\begin{equation}\label{eq:dist-local-decision-making-robustness-two-sided}
\begin{array}{cll}
 \displaystyle \min_{\vec x \in \cal X, l \ge 0} & l \\
 \st & |h(\vec x, \P_{\vec \xi}) - h(\vec x_0, \P_{0, \vec \xi})| \le l \cdot \epsilon, &\forall \P_{\vec \xi} \in \cal B_\epsilon(\P_{0, \vec \xi}), \\
 & h(\vec x_0, \P_{0, \vec \xi}) + l \cdot \epsilon \le t.
\end{array}
\end{equation}
Suppose that the cost threshold $t$ is sufficiently large so that the corresponding constraints can be dropped. Problems \eqref{eq:decision-making-robustness-two-sided} and \eqref{eq:dist-local-decision-making-robustness-two-sided} can be transformed into \textit{two-sided robust optimization}
\begin{equation}
\min_{\vec x \in \cal X} \max_{\vec \xi \in \cal B_\epsilon(\vec \xi_0)} |h(\vec x, \vec \xi) - h(\vec x_0, \vec \xi_0)|
\end{equation}
and \textit{two-sided distributionally robust optimization}
\begin{equation}
\min_{\vec x \in \cal X} \max_{\P_{\vec \xi} \in \cal B_\epsilon(\P_{0, \vec \xi})} |h(\vec x, \P_{\vec \xi}) - h(\vec x_0, \P_{0, \vec \xi})|,
\end{equation}
respectively. For example, in robust statistics, we aim to solve
\begin{equation}\label{eq:robust-staitics-min-max}
  \min_{T^*(\cdot)} \qquad \max_{\P_{0,\vec \xi} \in \cal F,~\P_{\vec \xi}: d(\P_{\vec \xi}, \P_{0,\vec \xi}) \le \epsilon}  \|T^*(\P_{\vec \xi}) - T(\P_{0,\vec \xi})\|,
\end{equation}
where $T^*$ denotes a robust estimator; compare with \eqref{eq:robust-statistics-spirit} where the quantity of interest is $T(\P_{0, \vec \xi})$ but $T$ itself may not be inherently robust. The maximization over $\P_{0,\vec \xi} \in \cal F \subseteq \cal M$ is required because $\P_{0,\vec \xi}$ is unknown but arbitrary on $\cal F$. Recall that the mathematical nature of robustness is the Lipschitz continuity of the cost function due to the arbitrariness of $\P_{0,\vec \xi}$. Formulation \eqref{eq:robust-staitics-min-max} reflects the spirit of \textit{robust estimation} \cite{zhu2022generalized,loh2024theoretical,liu2023robust}, which will be revisited later. Likewise, one-sided robust decision-making can be formulated as 
\begin{equation}\label{eq:decision-making-robustness-one-sided}
\begin{array}{cll}
 \displaystyle \min_{\vec x \in \cal X, l \ge 0} & l \\
 \st & h(\vec x, \vec \xi) - h(\vec x_0, \vec \xi_0) \le l \cdot \epsilon, &\forall \vec \xi \in \cal B_\epsilon(\vec \xi_0), \\
 & h(\vec x_0, \vec \xi_0) + l \cdot \epsilon \le t,
\end{array}
\end{equation}
while one-sided distributionally robust decision-making can be stated as
\begin{equation}\label{eq:dist-local-decision-making-robustness-one-sided}
\begin{array}{cll}
 \displaystyle \min_{\vec x \in \cal X, l \ge 0} & l \\
 \st & h(\vec x, \P_{\vec \xi}) - h(\vec x_0, \P_{0, \vec \xi}) \le l \cdot \epsilon, &\forall \P_{\vec \xi} \in \cal B_\epsilon(\P_{0, \vec \xi}), \\
 & h(\vec x_0, \P_{0, \vec \xi}) + l \cdot \epsilon \le t.
\end{array}
\end{equation}
Problems \eqref{eq:decision-making-robustness-one-sided} and \eqref{eq:dist-local-decision-making-robustness-one-sided} can be reformulated to validate the grounds of min-max \textit{optimization}.

\begin{highlight}[Min-Max Robustness]
By eliminating variables $l$, in the sense of the same robust solution(s) $\vec x^*$, \eqref{eq:decision-making-robustness-one-sided} is equivalent to the \textit{min-max robust optimization}
\begin{equation}\label{eq:decision-making-robustness-min-max-true}
    \min_{\vec x \in \cal X} \max_{\vec \xi \in \cal B_\epsilon(\vec \xi_0)} h(\vec x, \vec \xi),
\end{equation}
and \eqref{eq:dist-local-decision-making-robustness-one-sided} is equivalent to the \textit{min-max distributionally robust optimization}
\begin{equation}\label{eq:decision-making-dist-robustness-min-max-true}
    \min_{\vec x \in \cal X} \max_{\P_{\vec \xi} \in \cal B_\epsilon(\P_{0, \vec \xi})} h(\vec x, \P_{\vec \xi}),
\end{equation}
provided that $t$ is sufficiently large to guarantee the $t$-upper boundedness of the cost functions. \stp
\end{highlight}

Let $(\vec x^*, \vec \xi^*)$ denote the worst-case optimal decision and worst-case realization, respectively, that solve \eqref{eq:decision-making-robustness-min-max-true}. Problem \eqref{eq:decision-making-robustness-min-max-true} indicates that if the prescribed performance threshold $t$ is no smaller than the worst-case optimal cost $h(\vec x^*, \vec \xi^*)$, then the min-max formulation \eqref{eq:decision-making-robustness-min-max-true} computationally finds the robust solution $\vec x^*$ to \eqref{eq:decision-making-robustness-one-sided}, with one-sided local robustness measure $l^* \defeq [h(\vec x^*, \vec \xi^*) - h(\vec x_0, \vec \xi_0)]/\epsilon$. The similar statements can be straightforwardly extended for the distributional robustness case between \eqref{eq:dist-local-decision-making-robustness-one-sided} and \eqref{eq:decision-making-dist-robustness-min-max-true}: Letting $(\vec x^*, \P^*_{\vec \xi})$ denote the worst-case optimal decision and worst-case distribution, respectively, the one-sided local distributional robustness measure can be calculated as $l^* \defeq [h(\vec x^*, \P^*_{\vec \xi}) - h(\vec x_0, \P_{0,\vec \xi})]/\epsilon$. Problems \eqref{eq:decision-making-robustness-min-max-true} and \eqref{eq:decision-making-dist-robustness-min-max-true} justify the rationale of min-max (distributionally) robust optimization models that have been widely used in practice across diverse research areas: Namely, the real-world performance of a robust decision $\vec x^*$ can be protected for all perturbed parameters (resp. distributions) that are close to $\vec \xi_0$ (resp. $\P_{0, \vec \xi}$). In practice, however, $\vec \xi_0$ is unknown, so \eqref{eq:decision-making-robustness-min-max-true} cannot be directly solved. The same dilemma occurs in \eqref{eq:decision-making-dist-robustness-min-max-true}. Therefore, we have the following operational trick.
\begin{highlight}[Min-Max Robustness in Practice]
Suppose that $\vech \xi \in \cal B_\epsilon(\vec \xi_0)$ is perturbed from $\vec \xi_0$ and is known. We can approximately attack \eqref{eq:decision-making-robustness-min-max-true} resorting to
\begin{equation}\label{eq:decision-making-robustness-min-max}
    \min_{\vec x \in \cal X} \max_{\vec \xi \in \cal B_{\epsilon}(\vech \xi)} h(\vec x, \vec \xi),
\end{equation}
because if $\vech \xi \in \cal B_\epsilon(\vec \xi_0)$, we have $\vec \xi_0 \in \cal B_{\epsilon}(\vech \xi)$. 
The same operation applies to \eqref{eq:decision-making-dist-robustness-min-max-true}, leading to
\begin{equation}\label{eq:decision-making-dist-robustness-min-max}
    \min_{\vec x \in \cal X} \max_{\P_{\vec \xi} \in \cal B_{\epsilon}(\Ph_{\vec \xi})} h(\vec x, \P_{\vec \xi}),
\end{equation}
where $\Ph_{\vec \xi}$ is a perturbation of $\P_{0, \vec \xi}$.
\stp
\end{highlight}

In addition to the above change-of-center trick, motivated by \eqref{eq:robust-staitics-min-max}, another strategy is to simultaneously maximize \eqref{eq:decision-making-robustness-min-max-true} over $\vec \xi_0$ and \eqref{eq:decision-making-dist-robustness-min-max-true} over $\P_{0, \vec \xi}$ on some specified regions. However, in the practice of robust decision making, the change-of-center trick is the most popular choice; see, e.g., \cite{ben2009robust,kuhn2025distributionally}.

After formalizing the robustness principles and concepts, we next demonstrate how they can be used in designing robust statistical, optimization, and machine learning methods for diverse WSC (and general engineering) problems under uncertainties. To close this subsection, we compare the decision principle of robustness with that of adaptivity.

\begin{highlight}[Robustness Versus Adaptivity]
In adaptive design, the technical focus is placed on refining the estimate $\vech \xi$ of the true parameter $\vec \xi_0$ and then finding the adaptive solution $\vech x$ that minimizes $h(\vec x, \vech \xi)$. This approach relies on the assumption that $\vech \xi$ is a sufficiently accurate estimate of $\vec \xi_0$. In contrast, robust design explicitly acknowledges the potential estimation error in $\vech \xi$ and seeks a shared decision $\vec x^*$ that ensures satisfactory performance for all $\vech \xi$ within a reasonable neighborhood of $\vec \xi_0$. To clarify further, in the time-varying case, whenever the true parameter $\vec \xi_0$ changes, the adaptive strategy updates its estimate $\vech \xi$ and solution $\vech x$ in response to this change. In contrast, in robust design, even if $\vec \xi_0$ varies, as long as an estimate $\vech \xi$ remains within a reasonable neighborhood of $\vec \xi_0$, the estimate $\vech \xi$ and robust solution $\vec x^*$ are retained. Analogous statements regarding distributional robustness and adaptivity with respect to $\Ph_{\vec \xi}$, $\P_{0,\vec \xi}$, and $h(\vec x, \P_{\vec \xi})$ can be readily established, and hence, they are omitted. 
\stp
\end{highlight}

\subsection{Methodological Frameworks of Robustness}
In this subsection, we discuss modeling and computational frameworks for achieving robustness under uncertainties, leveraging the mathematical formalism of robustness concepts and principles. Specifically, we cover robust statistics, robust and distributionally robust optimization, and robust machine learning. These three technical areas collectively define the core scope of modern robust signal processing, encompassing both model-driven (i.e., physics-informed) and data-driven approaches. Model-driven methods rely on underlying physical laws, the mechanisms of data-generating processes, and domain-specific prior knowledge to build mathematical models. All subsequent analysis, computation, and decision-making are carried out based on these models. In contrast, data-driven methods depend primarily on historical data to uncover hidden patterns, extract new knowledge, approximate unknown functions, and make empirical (i.e., data-centric) decisions. In this sense, traditional forms of statistics and optimization are typically model-driven, whereas machine learning is largely data-driven, although the boundary between the two paradigms is increasingly blurred in modern practice. To clarify further, traditional statistics assumes a specific data-generating process (e.g., a parametric probabilistic distribution), and aims to infer population characteristics from observational data. Traditional optimization seeks to find the best solution to an objective function under given constraints, where both the objective and constraints are explicitly defined. In comparison, machine learning typically imposes no or few assumptions about the structural or distributional characteristics of the underlying data-generating process, and discovers patterns (e.g., clustering) or predictive mappings (e.g., classifiers) directly from collected data. Nevertheless, note that machine learning still relies on implicit representational and algorithmic assumptions to conduct the learning procedure, such as the choice of hypothesis spaces (e.g., linear functions, Lipschitz functions, convolutional neural networks, transformers, kernel embeddings), loss functions, and regularization strategies. The boundary blurring between statistics and machine learning arises primarily from the development of non-parametric statistics (e.g., Gaussian process regression), whereas that between optimization and machine learning stems from the emergence of data-driven and learning-assisted optimization methods (e.g., empirical risk minimization, algorithmic unrolling). To serve the pedagogical purpose, the level of technical detail is carefully balanced in this subsection, with greater emphasis placed on motivations, concepts, interpretations, and illustrations; provided references can guide readers to further developments.


\subsubsection{Robust Statistics}
Statistics is concerned with analyzing, interpreting, and inferring the characteristics of a dataset and its population, for example, investigating the data-generating process (e.g., linear or nonlinear), identifying the underlying probabilistic distribution (e.g., Gaussian versus Laplacian), quantifying measures of location and dispersion (e.g., mean, variance), and examining relationships among subgroups or between covariates (e.g., causal or noncausal). Representative applications of statistics in WCS engineering include channel modeling (e.g., Rayleigh or Rician fading), channel estimation (e.g., maximum likelihood or Bayesian estimation), transmitted-signal detection (e.g., on a constellation), target detection (e.g., matched filtering), target tracking (e.g., Kalman filtering), direction-of-arrival estimation (e.g., maximum likelihood methods), to name a few. 

Statistical inference is well-known to be vulnerable to even slight modeling mismatches due to insufficiency of physical knowledge, non-adversarial distributional shifts due to environmental variations or input-condition changes (e.g., sensor biases), adversarial data corruptions due to data poisoning attacks (e.g., outliers), etc \cite{loh2024theoretical,huber2009robust,zhu2022generalized}. In this subsection, we particularly discuss robust estimation, hypothesis testing, and regression techniques that can perform well under these uncertainties; for introductory surveys, see \cite{loh2024theoretical,diakonikolas2019robust}. Robust estimation of mean and covariance plays a critical role in channel estimation in wireless communications (WC) and adaptive beamforming in wireless sensing (WS). Robust hypothesis testing is essential for channel change detection (WC) and radar target detection (WS), while robust regression is fundamental to transmitted-signal detection (WC) and target localization (WS).

Formal studies of robust statistics began in the 1960s, for example, robust mean and scale estimation, robust testing, and robust regression to fight against outliers (e.g., impulse noises). A representative tutorial-style overview article in signal processing that echoes these robust statistical techniques is \cite{zoubir2012robust}. However, these works primarily addressed univariate problems, and their robust solutions do not directly extend to high-dimensional settings, because computational complexity grows exponentially with dimensionality \cite{diakonikolas2019robust,diakonikolas2021robustness}. High-dimensional statistical problems are common in WSC systems that employ MIMO antenna arrays. Only since around 2016 have computationally efficient methods for high-dimensional robust statistics emerged \cite{lai2016agnostic}. This article, therefore, also covers recent advances in high-dimensional robust statistics to motivate and guide signal processing applications.

To perform a statistical procedure, two fundamental questions are immediate: First, what are the characteristics of the population distribution? Second, what quantity to estimate or hypothesis to test? In parametric statistics, the population is characterized by a parametric probabilistic distribution $\P_{\vec \xi}(\vec \theta)$, e.g., a Gaussian or exponential, where $\vec \theta$ denotes the defining parameters (e.g., mean, covariance, median) of the distribution $\P_{\vec \xi}$. In this case, the quantity to estimate is natural, i.e., $\vec \theta$, so are the hypotheses to test, i.e., $\P_{\vec \xi}(\vec \theta_1)$ versus $\P_{\vec \xi}(\vec \theta_2)$. However, in non-parametric statistics, the population distribution $\P_{\vec \xi}$ is not parameterized by any real-valued number(s). Instead, $\P_{\vec \xi}$ may just be stated as \quotemark{a distribution that is symmetric about zero, has a continuous density function, and has finite first two moments}. In this case, the quantity to estimate and the hypotheses to test need to be specified in accordance with the problem of interest. In robust statistics, two more questions arise: What uncertainties are a robust statistical procedure expected to combat (e.g., outliers in observations), and how to quantify these uncertainties relative to the clean data distribution? 

We begin with robust estimation. 
Let $T(\P_{\vec \xi})$ be a statistical functional, called a \textit{descriptive measure} of $\P_{\vec \xi}$, that specifies the quantity to be estimated \cite[pp.~11-15]{staudte1990robust}. For example, in the parametric case, $T(\P_{\vec \xi}(\vec \theta))$ returns $\vec \theta$; supposing $\P_{\xi}(\mu, 1)$ is a Gaussian distribution with mean $\mu$ and variance $1$, whose density function is $\varphi(\xi; \mu, 1)$, we can define 
\begin{equation}\label{eq:mean-estimator}
    T(\P_{\xi}(\mu, 1)) \defeq \mu = \int \xi \varphi(\xi; \mu, 1) \d \xi
\end{equation}
for univariate mean estimation. For another example, $T(\P_{\vec \xi})$ can denote the location (e.g., coordinate-wise median) or scatter (e.g., covariance) of a high-dimensional non-parametric distribution $\P_{\vec \xi}$. If $\Ph_{n, \vec \xi}$ denotes an empirical distribution constructed using $n$ samples drawn from $\P_{\vec \xi}$, the statistical functional $T$ can naturally induce an estimator $T(\Ph_{n, \vec \xi})$ of $T(\P_{\vec \xi})$, and under some technical conditions (e.g., $T$ is weakly continuous \cite[p.~42]{huber2009robust}), we have 
$
T(\Ph_{n, \vec \xi}) \to T(\P_{\vec \xi})
$ 
as $n \to \infty$; recall the law of large numbers. Typical instances include mean and covariance estimation: e.g., $\sum^n_{i=1} \vec \xi_i / n \to \E_{\P_{\vec \xi}} \vec \xi$. However, when the operating data-generating distribution $\P_{\vec \xi}$ deviates from the interested clean data distribution $\P_{0, \vec \xi}$, $T(\Ph_{n, \vec \xi})$ may not be an reliable estimate of $T(\P_{0,\vec \xi})$. This can be seen from 
\begin{equation}
    \| T(\Ph_{n, \vec \xi}) - T(\P_{0,\vec \xi}) \| \le \| T(\Ph_{n, \vec \xi}) - T(\P_{\vec \xi}) \| + \| T(\P_{\vec \xi}) - T(\P_{0,\vec \xi}) \|,
\end{equation}
where the first term $\| T(\Ph_{n, \vec \xi}) - T(\P_{\vec \xi}) \|$ on the right-hand side can vanish as the data size $n$ tends to infinity, while the second term $\| T(\P_{\vec \xi}) - T(\P_{0,\vec \xi}) \|$ can be uncontrollable; $\|\cdot\|$ denotes an appropriate vector norm. Therefore, robust estimation naturally requires the statistical functional $T$ to conform to the following \textit{Principle of Distributional Robustness}, as defined previously:
\begin{equation}\label{eq:robust-estimation-principle}
    d(\P_{\vec \xi}, \P_{0,\vec \xi}) \le \epsilon \implies \| T(\P_{\vec \xi}) - T(\P_{0,\vec \xi}) \| \le l \epsilon,
\end{equation}
for some finite, or preferably small, values of $l$. For estimation, the absolute closeness between $T(\Ph_{n, \vec \xi})$ and $T(\P_{0,\vec \xi})$ matters, so the two-sided (or all-directional) distributional robustness of an object applies to robust estimation; recall \eqref{eq:robust-statistics-spirit}. When $T$ is not inherently distributionally robust against the non-zero distributional shift $d(\P_{\vec \xi}, \P_{0,\vec \xi}) \le \epsilon$, a robust counterpart $T^*$ of $T$, called a \textit{robust estimator}, is demanded to ensure
\begin{equation}\label{eq:principle-robust-estimation}
    \| T^*(\Ph_{n, \vec \xi}) - T(\P_{0,\vec \xi}) \| \to  \| T^*(\P_{ \vec \xi}) - T(\P_{0,\vec \xi}) \| \le l \epsilon,
\end{equation}
as $n \to \infty$, for some finite $l$; the smaller the value of $l$, the more robust the estimator $T^*$. For example, the mean estimator in \eqref{eq:mean-estimator} is non-robust against large outliers. Given $\epsilon \in (0, 1)$, let the clean data distribution $\P_{0,\xi}$ have the standard Gaussian density $\varphi(\xi; 0, 1)$ and the operating data-generating distribution $\P_{\xi}$ have the Hampel's $\epsilon$-mixture density \cite{hampel1974influence}
\begin{equation}\label{eq:huber-dist}
    (1-\epsilon) \varphi(\xi; 0, 1) + \epsilon \delta_{\xi^\prime}(\xi);
\end{equation}
that is, with probability $1-\epsilon$, a sample is drawn from the clean data distribution $\P_{0, \xi}$, and with probability $\epsilon$, a sample is from a Dirac distribution $\delta_{\xi^\prime}(\xi)$ concentrated at the point $\xi^\prime$. Hence, although the total variation (TV) distance between $\P_{\xi}$ and $\P_{0, \xi}$ is no larger than $\epsilon$, i.e.,
\begin{equation}\label{eq:TV-Hampel}
d_{\text{TV}}(\P_{\xi}, \P_{0,\xi}) = \frac{1}{2}\int^{\infty}_{-\infty} \Big|-\epsilon \varphi(\xi; 0, 1) + \epsilon \delta_{\xi^\prime}(\xi)\Big| \d \xi \le \epsilon,
\end{equation} 
the absolute deviation between the mean of $\P_{\xi}$ and that of $\P_{0, \xi}$ equals
\[
| T(\P_{\xi}) - T(\P_{0, \xi}) | = |(1 - \epsilon) \cdot 0+ \epsilon \xi^\prime - 0| = \epsilon \xi^\prime,
\]
which can be arbitrarily large without a finite upper bound, due to the arbitrariness of the outlier $\xi^\prime$. However, in this case, given $\epsilon < 0.5$, the median (i.e., $50\%$-quintile) estimator $T^*(\P_{\xi})$ is robust \cite[Section~2]{chen2018robust}: that is, there exists a finite $l$ such that
\[
| T^*(\P_{\xi}) - T(\P_{0, \xi}) | \le l \epsilon.
\]
For Gaussian clean distributions, empirical estimators of covariances (i.e., sample covariances) are also extremely sensitive to even a single large outlier in the data. 

As indicated, the mathematical nature of robust estimation can be stated as follows.
\begin{highlight}[Robust Estimation]
Given the uncertainty quantification $d(\P_{\vec \xi}, \P_{0,\vec \xi}) \le \epsilon$ under an appropriate similarity measure $d$, 
a sample-based robust estimate $T^*(\Ph_{n, \vec \xi})$ of the clean-distribution quantity $T(\P_{0,\vec \xi})$ should satisfy
\begin{equation}\label{eq:rule-robust-estimation}
\begin{array}{l}
\| T^*(\Ph_{n, \vec \xi}) - T(\P_{0,\vec \xi}) \| \\
\quad \quad \le \| T^*(\Ph_{n, \vec \xi}) - T^*(\P_{\vec \xi}) \| + \| T^*(\P_{\vec \xi}) - T(\P_{0,\vec \xi}) \| \\
\quad \quad \le g(n) + l \epsilon \\
\quad \quad \to l \epsilon,
\end{array}
\end{equation}
in probability or expectation (due to the randomness of $\Ph_{n, \vec \xi}$), for some constants $l < \infty$ and convergence-rate functions $g(n) \ge 0$, where $g(n)$ characterizes the finite-sample effect and approaches zero as $n \to \infty$.
\stp
\end{highlight} 

The above principle of robustness is widely adopted in modern robust statistics. In some robust mean and covariance estimation cases, $g(n) = c\sqrt{m/n}$ for some constants $c$, where $m$ is the dimension of $\vec \xi$. For technical details, see \cite{lai2016agnostic,diakonikolas2019robust,chen2018robust,zhu2022generalized,gao2019robust,liu2023robust}. Just note that although the linear robustness upper bound $l\epsilon$ can be information-theoretically guaranteed \cite{chen2018robust,gao2019robust,liu2023robust}, some robust estimators (e.g., \cite{lai2016agnostic,diakonikolas2019robust}) can only computationally achieve a looser error bound $e(\epsilon) \ge \epsilon$ such that $\| T^*(\P_{\vec \xi}) - T(\P_{0,\vec \xi}) \| \le l e(\epsilon)$ and $e(\epsilon) \to 0$ as $\epsilon \to 0$. This is a well-known tradeoff in robust estimation: the linear robustness property, which is information-theoretically optimal \cite{chen2018robust,liu2023robust}, might be compromised to achieve computational efficiency.

\begin{figure}[!htbp]
	\centering
	\subfigure[TV Distance]{
	 	\includegraphics[width=4cm]{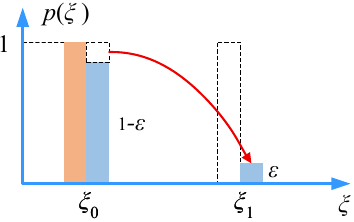}
	}~~~
	\subfigure[Wasserstein Distance]{
	 	\includegraphics[width=4cm]{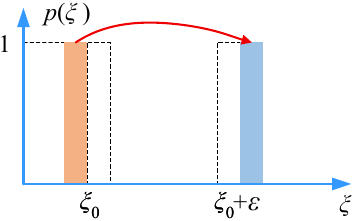}
	}
    
	\caption{Comparison between the TV and Wasserstein distances: the former cannot capture the underlying geometry between points, while the latter can. In (a), two distributions are supported on $\xi_0$ and $\xi_1$, with probability mass vectors $p_0 = [1, 0]$ and $p_1 = [1-\epsilon, \epsilon]$, respectively. In this case, the TV distance $d_{\text{TV}}(p_0, p_1) = \epsilon$, no matter how large $|\xi_1 - \xi_0|$ is. In contrast, the Wasserstein distance $d_{\text{W}}(p_0, p_1) = \epsilon \cdot |\xi_1 - \xi_0|$ can be infinite if $|\xi_1 - \xi_0| \to \infty$. In (b), we let $\xi_1 = \xi_0 + \epsilon$ for $\epsilon > 0$, $p_0 = [1, 0]$, and $p_1 = [0, 1]$. In this case, $d_{\text{TV}}(p_0, p_1) = 1$, no matter how small $\epsilon$ is. However, $d_{\text{W}}(p_0, p_1) = \epsilon$.}
	\label{fig:TV-W}
\end{figure}

In the practice of applied statistics, operations research, and machine learning, two common types of data corruption or contamination exist \cite{zhu2022generalized,liu2023robust,loh2024theoretical}. In the first type, called partial contamination, only a subset of the clean data is contaminated, e.g., replaced with arbitrary values, while the remaining samples remain unchanged. Unexpected outliers in the data can be probabilistically modeled in this way, and the Huber's $\epsilon$-contamination model
\begin{equation}\label{eq:huber-contamination-model}
    \P_{\vec \xi} \defeq (1-\epsilon)\P_{0, \vec \xi} + \epsilon \Q
\end{equation}
is standard, where $\Q$ denotes any contamination distribution \cite{huber2009robust,chen2018robust}; recall a special case in \eqref{eq:huber-dist}. For this type of contamination, the total variation distance is employed to characterize the distributional shift \cite{zhu2022generalized,diakonikolas2019robust,gao2019robust}, i.e., 
\begin{equation}\label{eq:TV-distance}
d_{\text{TV}}(\P_{\vec \xi}, \P_{0,\vec \xi}) \defeq \inf_{\pi \in \Pi(\P_{\vec \xi}, \P_{0, \vec \xi})} \int_\pi \bb I_{\vec \xi \ne \vec \xi_0} \d \pi(\vec \xi, \vec \xi_0) \le \epsilon,
\end{equation}
where $\Pi(\P_{\vec \xi}, \P_{0, \vec \xi})$ contains all joint distributions of $(\vec \xi, \vec \xi_0)$ that have marginals $\P_{\vec \xi}$ and $\P_{0, \vec \xi}$, respectively; $\bb I_{\vec \xi \ne \vec \xi_0}$ equals one if $\vec \xi \ne \vec \xi_0$ and zero otherwise. 
Namely, if $d_{\text{TV}}(\P_{\vec \xi}, \P_{0,\vec \xi}) = 0$, it must have $\P_{\vec \xi} = \P_{0, \vec \xi}$, and vise versa. Outlier-robustness research in this stream includes classical methods for one-dimensional estimation problems \cite{hampel1974influence,huber2009robust,staudte1990robust} and recent computationally efficient algorithms for high-dimensional ones \cite{lai2016agnostic,diakonikolas2019robust}. In the second type, called full contamination, all clean data samples can be perturbed. Measurement biases in sensors can introduce such small perturbations across all samples. For this type of contamination, the $p$-Wasserstein distance can be used to quantify the distributional shift \cite{zhu2022generalized,liu2023robust,nguyen2022distributionally}, i.e., 
\begin{equation}\label{eq:W-distance}
d_{\text{W},p}(\P_{\vec \xi}, \P_{0, \vec \xi}) \defeq \left[\inf_{\pi \in \Pi(\P_{\vec \xi}, \P_{0, \vec \xi})} \int_\pi c^p(\vec \xi, \vec \xi_0) \d \pi(\vec \xi, \vec \xi_0) \right]^{1/p} \le \epsilon,
\end{equation}
where $c$ is a generic distance (other than $\bb I_{\vec \xi \ne \vec \xi_0}$) and $\P_{\vec \xi}$ denotes the distribution of the perturbed data; $p$ is notationally dropped if it equals one. Note that when $d_{\text{TV}}(\P_{\vec \xi}, \P_{0,\vec \xi})$ is tiny, the Wasserstein distance $d_{\text{W}, p}(\P_{\vec \xi}, \P_{0,\vec \xi})$ can be infinite; when $d_{\text{W}, p}(\P_{\vec \xi}, \P_{0, \vec \xi})$ is tiny, the total variation distance $d_{\text{TV}}(\P_{\vec \xi}, \P_{0,\vec \xi})$ can reach its natural upper bound, i.e., one. This is because the total variation distance ignores the underlying geometry between $\vec \xi$ and $\vec \xi_0$, as it only uses a binary cost function $\bb I_{\vec \xi \ne \vec \xi_0}$ to distinguish whether two points are equal or not. In contrast, the Wasserstein distance incorporates the geometry by using a generic cost function $c(\vec \xi, \vec \xi_0)$ that depends on the actual closeness between $\vec \xi$ and $\vec \xi_0$. Therefore, for the two contamination cases, the uncertainty quantification metrics cannot be shared. For more technical remarks on $d_{\text{TV}}$ and $d_{\text{W},p}$ in robust statistics, see \cite{zhu2022generalized,gao2019robust,liu2023robust} and references therein. For a visual demonstration, see Fig. \ref{fig:TV-W}; Fig. \ref{fig:TV-W}(a) straightforwardly explains why the TV distance is a proper choice to quantify outliers. 

In outlier-robust estimation based on the contamination model \eqref{eq:huber-contamination-model} for $\Q \defeq \delta_{\vec \xi^\prime}(\vec \xi)$ and an arbitrary point $\vec \xi^\prime$, the concept of \textit{influence function} (IF) is well studied to measure the influence of $\vec \xi^\prime$ on an estimator $T(\P_{\vec \xi})$, even when the amount $\epsilon$ of contamination is infinitesimal \cite{hampel1968contributions,hampel1974influence}. The influence of $\vec \xi^\prime$ on $T$ at $\P_{0, \vec \xi}$ is defined as
\begin{equation}\label{eq:influence-function}
    \operatorname{IF}(\vec \xi^\prime; T, \P_{0, \vec \xi}) \defeq \lim_{\epsilon \to 0}\frac{\| T[(1-\epsilon)\P_{0, \vec \xi} + \epsilon \delta_{\vec \xi^\prime}(\vec \xi)] - T[\P_{0,\vec \xi}] \|}{\epsilon}.
\end{equation}
The above definition of IF was originally proposed by Frank Hampel for both one-dimensional and high-dimensional estimation problems \cite[p.~37]{hampel1968contributions}. To achieve outlier-robustness, it requires
\[
l \defeq \sup_{\vec \xi^\prime \in \Xi} \operatorname{IF}(\vec \xi^\prime; T, \P_{0, \vec \xi}) < \infty.
\]
This uniform boundedness property of $\operatorname{IF}(\vec \xi^\prime; T, \P_{0, \vec \xi})$ on $\Xi$ naturally implies the distributional robustness \eqref{eq:robust-estimation-principle} of the estimator $T$ against the outlier distribution $\delta_{\vec \xi^\prime}(\vec \xi)$. Note that a functional with a bounded derivative is Lipschitz continuous. For a one-dimensional visual demonstration, see \cite[Fig.~2]{zoubir2012robust}. 

For both partial and full contamination types, another concept, called \textit{break-down point} (BP) \cite[p.~27]{hampel1968contributions}, is informative, which is defined as 
\begin{equation}\label{eq:breakdown-point}
    \operatorname{BP}(T; d, \P_{0, \vec \xi}) \defeq \sup_{\epsilon \ge 0} \left\{\epsilon \left| 
    \begin{array}{l}
         \| T(\P_{\vec \xi}) - T(\P_{0,\vec \xi}) \| < \infty \\
         d(\P_{\vec \xi}, \P_{0,\vec \xi}) \le \epsilon
    \end{array}
    \right.\right\},
\end{equation}
where $d$ can be either $d_{\text{TV}}$ for partial contamination or $d_{\text{W},p}$ for full contamination. Therefore, in real-world operation, an estimator $T$ is robust whenever the distributional shift satisfies 
\[
d(\P_{\vec \xi}, \P_{0,\vec \xi}) \le \operatorname{BP}(T; d, \P_{0, \vec \xi}).
\]
As indicated, the BP property is ad hoc to a specified contamination type and its quantification. For a one-dimensional visualization in the outlier-robust case, see \cite[Fig.~1]{zoubir2012robust}.

One standard robust estimation framework is M-estimation, which was first introduced by Peter Huber \cite{huber1964robsut}, \cite[Section~3.2]{huber2009robust}. M-estimation $T^*(\P_{\vec \xi})$ for an clean-data quantity $T(\P_{0, \vec \xi})$ is defined as
\begin{equation}\label{eq:M-estimation}
    T^*(\P_{\vec \xi}) \defeq \argmin_{\vec \theta} \E_{\P_{\vec \xi}} h(\vec \xi, \vec \theta),
\end{equation}
for some cost functions $h$. For location (e.g., mean) estimation, we have $h(\vec \xi, \vec \theta) \defeq h(\vec \xi - \vec \theta)$. Under some technical conditions of $h$ (e.g., the ${\vec \theta}$-derivative $\nabla_{\vec \theta} h(\vec \xi, \vec \theta)$ is bounded or Lipschitz continuous in $\vec \xi$), it holds that \cite{romisch2003stability}
\begin{equation}
\|T^*(\P_{\vec \xi}) - T(\P_{0, \vec \xi})\| \le l \cdot d(\P_{\vec \xi}, \P_{0, \vec \xi}),
\end{equation}
for some finite constants $l$. Here, $d$ can be either $d_{\text{TV}}$ for partial contamination (e.g., outliers) and bounded derivative, or $d_{\text{W},p}$ for full contamination (e.g., sensor biases) and Lipschitz continuous derivative. Therefore, to obtain robust M-estimators, one can begin by designing suitable cost functions $h$. For example, under the Hampel's $\epsilon$-contamination model \eqref{eq:huber-dist}, the influence function $\operatorname{IF}(\vec \xi)$ of the M-estimator $T^*$ in \eqref{eq:M-estimation} is proportional to the derivative $\psi(\vec \xi, \vec \theta_0) \defeq \nabla_{\vec \theta} h(\vec \xi, \vec \theta) |_{\vec \theta = \vec \theta_0}$ \cite[p.~387]{hampel1974influence}, \cite[Eq.~(3.13)]{huber2009robust}. For this reason, to design robust estimators against outliers, one can begin by designing bounded influence functions $\operatorname{IF}(\vec \xi)$ \cite{hampel1974influence,zoubir2012robust,loh2024theoretical} such that $\psi(\vec \xi, \vec \theta)$ can be bounded in $\vec \xi$ for every $\vec \theta$. Consequently, by minimizing \eqref{eq:M-estimation} with the induced cost function $h(\vec \xi, \vec \theta)$, outlier-robust estimators can be obtained. To be more specific, in one-dimensional mean estimation $T^*(\P_{\xi}) \defeq \argmin_{\theta} \E_{\P_{\xi}} h(\xi - \theta)$ with $\P_{\xi}$ given in \eqref{eq:huber-dist}, the quadratic function $h(z) \defeq z^2$ leads to the outlier-non-robust arithmetic mean estimator, the absolute function $h(z) \defeq |z|$ results in the outlier-robust median estimator, and the Huber's cost function 
\begin{equation}\label{eq:Lhuber-cost}
    h(z) = \left\{
    \begin{array}{ll}
        \frac{1}{2} z^2,  &  \text{if } |z| \le k,\\
        k |z| - \frac{1}{2} k^2, & \text{otherwise}
    \end{array}
    \right.
\end{equation}
yields the outlier-robust Huber estimator, where $k$ is a constant determined by $\epsilon$ \cite{huber1964robsut}. As indicated, the $h$-manipulating procedure captures the essence of robust M-estimation that combats either partial contamination (e.g., outliers) under $d_{\text{TV}}$ [see \eqref{eq:TV-Hampel}] or full contamination (e.g., sensor biases) under $d_{\text{W},p}$; the former requires a bounded derivative of $h$, while the latter demands a Lipschitz continuous derivative. 

For mean and covariance estimation under the Gaussian clean-data distribution $\P_{0, \vec \xi}$, the problem is formulated as
\begin{equation}\label{eq:estimation-mean-cov}
T(\P_{0, \vec \xi}) \defeq \displaystyle \argmin_{\vec \mu, \mat \Sigma \succeq \mat 0} \E_{\P_{0, \vec \xi}} 
(\vec \xi - \vec{\mu})^{\top}
\mat{\Sigma}^{-1}
(\vec \xi - \vec{\mu})
+ \log \det \mat{\Sigma}.
\end{equation}
One can verify that the problem is convex and smooth in $\vec \mu$ and $\mat \Sigma$, so the first-order optimality conditions yield the optimal estimates, i.e., $\vec \mu_0 = \E_{\P_{0,\vec \xi}} \vec \xi$ and $\mat \Sigma_0 = \E_{\P_{0,\vec \xi}} (\vec \xi - \vec \mu_0)(\vec \xi - \vec \mu_0)^\T$. However, whenever the clean data distribution $\P_{0, \vec \xi}$ is contaminated into $\P_{\vec \xi}$, either partially or fully, robustification of the estimator \eqref{eq:estimation-mean-cov} is required.

For robust estimation of mean and covariance under partial contamination (i.e., outliers), to achieve bounded influences of outliers on the estimator, M-estimation \eqref{eq:M-estimation} particularizes into 
\begin{equation}\label{eq:M-estimation-mean-cov}
T^*(\P_{\vec \xi}) \defeq \displaystyle \argmin_{\vec \mu, \mat \Sigma \succeq \mat 0} \E_{\P_{\vec \xi}} 
\rho\!\left[
(\vec \xi - \vec{\mu})^{\top}
\mat{\Sigma}^{-1}
(\vec \xi - \vec{\mu})
\right]
+ \log \det \mat{\Sigma},
\end{equation}
where $\rho(\cdot)$ is a cost function that determines the influence of a sample on the estimator; recall from \eqref{eq:rule-robust-estimation} that in real-world operation, $\P_{\vec \xi}$ is replaced with its sample-based empirical estimate $\Ph_{n, \vec \xi}$. If $\rho(z) = z$, \eqref{eq:M-estimation-mean-cov} degenerates to the usual mean and covariance estimator \eqref{eq:estimation-mean-cov}, which is non-robust to outliers; if $\rho(z) = \d h(z)/\d z$ where $h(z)$ is defined in \eqref{eq:Lhuber-cost}, we have the Huber-type M-estimator that is outlier-robust due to the bounded impact of outliers. For a concrete example, see \cite{maronna1976robust}; note that the first-order optimality conditions of \eqref{eq:M-estimation-mean-cov} can be directly employed to obtain the solutions to \eqref{eq:M-estimation-mean-cov}. In addition to M-estimation, other approaches for robust location (e.g., mean) and dispersion (e.g., covariance) estimation against outliers include Tukey's median, geometric median, minimum volume enclosing ellipsoid, etc.; see \cite{lai2016agnostic,diakonikolas2021robustness,chen2018robust}. However, these proposals are criticized for their computational complexity, particularly in high-dimensional settings \cite{lai2016agnostic,diakonikolas2021robustness,diakonikolas2019robust}. Hence, computationally efficient algorithms have been developed since 2016. To date, most outlier-robust methods for mean $\vec \mu^*_{\text{TV}}$ and covariance $\mat \Sigma^*_{\text{TV}}$ estimation take the following weighted forms \cite{zoubir2012robust,maronna1976robust,maronna2002robust,diakonikolas2021robustness,lai2016agnostic}:
\begin{equation}\label{eq:weighted-robust-mean}
\vec \mu^*_{\text{TV}} \defeq  \frac{1}{n}\sum^n_{i = 1} w_i \cdot \vec \xi_i
\end{equation}
and
\begin{equation}\label{eq:weighted-robust-cov}
\mat \Sigma^*_{\text{TV}} \defeq \frac{1}{n}\sum^n_{i = 1} w_i \cdot (\vec \xi_i - \vec \mu^*_{\text{TV}})(\vec \xi_i - \vec \mu^*_{\text{TV}})^\T,
\end{equation}
where $w_i \ge 0$ denotes the weight of the sample $\vec \xi_i$. For non-robust cases as in \eqref{eq:M-estimation-mean-cov} for $\rho(z) = z$, we have $w_i = 1$. For robust cases, the weight $w_i$ may depend on $\vec \xi_i$, $\vec \mu^*_{\text{TV}}$, and $\mat \Sigma^*_{\text{TV}}$ (as in M-estimation), but this is not always the case (as in \cite{diakonikolas2021robustness,diakonikolas2019robust}). When $\vec \xi_i$ is identified as an outlier by a robust estimation algorithm, it has $w_i = 0$ for outlier rejection (cf. re-descending influence functions) or small $w_i$ for outlier damping (cf. monotonic influence functions) \cite{huber2009robust}. For such outlier-robust estimates via proper weighting, the robustness property \eqref{eq:rule-robust-estimation} for partial contamination under $d_{\text{TV}}$ can be established. 

For robust estimation of mean and covariance under full
contamination, estimators can be defined through worst-case optimization of \eqref{eq:estimation-mean-cov}. For example,
\begin{equation}\label{eq:wasserstein-mean-cov-estimation}
T^*(\P_{\vec \xi}) \defeq \displaystyle \argmin_{\vec \mu, \mat \Sigma \succeq \mat 0} \max_{\P \in \cal B_{\text{W}, 2, \epsilon}(\P_{\vec \xi})} \E_{\P} 
(\vec \xi - \vec{\mu})^{\top}
\mat{\Sigma}^{-1}
(\vec \xi - \vec{\mu})
+ \log \det \mat{\Sigma},
\end{equation}
where
\begin{equation}\label{eq:wasserstein-ball}
    \cal B_{\text{W}, 2, \epsilon}(\P_{\vec \xi}) \defeq \{ \P: d_{\text{W},2}(\P, \P_{\vec \xi}) \le \epsilon \}
\end{equation}
denotes the $2$-Wasserstein ball with center $\P_{\vec \xi}$ and radius $\epsilon$ \cite{nguyen2022distributionally}. In this case, we suppose that the clean data distribution $\P_{0, \vec \xi} $ is included in $\cal B_{\text{W}, 2, \epsilon}(\P_{\vec \xi})$, so the full contamination can be quantified by $d_{\text{W},2}(\P_{0, \vec \xi}, \P_{\vec \xi}) \le \epsilon$. After replacing $\P_{\vec \xi}$ with $\Ph_{n, \vec \xi}$, it can be proved that the robust mean estimate coincides with the empirical mean, i.e.,
\begin{equation}\label{eq:wasserstein-robust-mean}
\vec \mu^*_{\text{W}} \defeq \E_{\Ph_{n, \vec \xi}} \vec \xi = \frac{1}{n}\sum^n_{i = 1} \vec \xi_i,
\end{equation}
and \eqref{eq:wasserstein-mean-cov-estimation} reduces to the robust covariance estimation problem:
\begin{equation}\label{eq:wasserstein-cov-estimation}
\mat \Sigma^*_{\text{W}} \defeq \displaystyle \argmin_{\mat \Sigma \succeq \mat 0} \max_{\P \in \cal B_{\text{W}, 2, \epsilon}(\Ph_{n, \vec \xi})} \E_{\P} \langle \vec \xi \vec \xi^{\top},  \mat{\Sigma}^{-1} \rangle
+ \log \det \mat{\Sigma},
\end{equation}
where $\langle \mat \Sigma_1, \mat \Sigma_2 \rangle \defeq \Tr [\mat \Sigma^\top_1 \mat \Sigma_2]$ for two matrices \cite{nguyen2022distributionally}. Another similar min-max optimization-based framework for robust covariance estimation is established in \cite{yue2024geometric}. In essence, a robust covariance estimate shares the same eigen-vectors with the sample covariance $\math \Sigma_n$, and only eigen-values are manipulated: that is,
\begin{equation}\label{eq:eigen-manipulating}
    \mat \Sigma^*_{\text{W}} = \math V \mat X^* \math V^\T,
\end{equation}
where $\math V \math X \math V^\T$ denotes the eigen-decomposition of $\math \Sigma_n$, and the diagonal matrix $\mat X^*$ is modified from the empirical eigen-value matrix $\math X$; see \cite[Remark~3.2]{nguyen2022distributionally}, \cite[Theorem~1]{yue2024geometric}. 


The rationale of the robust estimation framework \eqref{eq:wasserstein-cov-estimation} can also be justified using the min-max optimization \eqref{eq:decision-making-dist-robustness-min-max}. To be specific, consider a loss function for covariance estimation as follows:
\[
h(\mat \Sigma, \mat \Sigma_0) \defeq \langle   \mat{\Sigma}_0, \mat \Sigma^{-1} \rangle
+ \log \det \mat{\Sigma} - \log \det \mat{\Sigma}_0  - m,
\]
where $m$ is the dimension of $\vec \xi$. When $\min_{\mat \Sigma \succeq \mat 0} h(\mat \Sigma, \mat \Sigma_0)$ is solved by $\mat \Sigma = \mat \Sigma_0$, the loss equals zero. Hence, the true optimization of interest is
\[
\min_{\mat \Sigma \succeq \mat 0} \langle   \E_{\P_{0, \vec \xi}} \vec \xi \vec \xi^\T, \mat \Sigma^{-1} \rangle
+ \log \det \mat{\Sigma}.
\]
For this cost-sensitive problem (i.e., the smaller the cost, the better), we achieve the one-sided robustness against all contaminated distributions in $\cal B_{\text{W}, 2, \epsilon}(\P_{0, \vec \xi})$ through \eqref{eq:dist-local-decision-making-robustness-one-sided}, which is algorithmically solved by \eqref{eq:decision-making-dist-robustness-min-max-true}, i.e., 
\begin{equation}\label{eq:viet-robust-cov}
\min_{\mat \Sigma \succeq \mat 0} \max_{\P \in \cal B_{\text{W}, 2, \epsilon}(\P_{0, \vec \xi})} \langle   \E_{\P} \vec \xi \vec \xi^\T, \mat \Sigma^{-1} \rangle
+ \log \det \mat{\Sigma}.
\end{equation}
Since the contaminated distribution $\P_{\vec \xi}$ lies in $\cal B_{\text{W}, 2, \epsilon}(\P_{0, \vec \xi})$, the above display can be approximately solved by the robust covariance estimation problem \eqref{eq:wasserstein-cov-estimation} after replacing $\P_{\vec \xi}$ with its empirical estimate $\Ph_{n, \vec \xi}$; recall the relationship between \eqref{eq:decision-making-dist-robustness-min-max-true} and \eqref{eq:decision-making-dist-robustness-min-max}.

The eigenvalue manipulating operation \eqref{eq:eigen-manipulating} is reminiscent of the linear shrinkage estimators of covariances that are more robust than the sample covariance $\math \Sigma_n$, which is a convex $\alpha$-combination of $\math \Sigma_n$ and the identity matrix $\mat I$ \cite{ledoit2004well,chen2010shrinkage}: i.e., 
\begin{equation}
\mat \Sigma^*_{\text{W}} = \alpha \math \Sigma_n + (1-\alpha) \mat I.
\end{equation}
The above linear shrinkage strategy, called diagonal loading \cite{mestre2005finite,wang2025distributionallybeamforming}, has been widely used in signal processing due to its computational simplicity compared to the complicated approaches in \cite{nguyen2022distributionally,yue2024geometric}. When some reliable prior information $\matb \Sigma$ of the true covariance $\mat \Sigma_0$ is accessible, robustness can also be achieved through prior-information embedding \cite{stoica2008using}
\begin{equation}
\mat \Sigma^*_{\text{W}} = \alpha \math \Sigma_n + (1-\alpha) \matb \Sigma.
\end{equation}
Indeed, when $\alpha$ is cleverly tuned, $\mat \Sigma^*_{\text{W}}$ can be closer to $\mat \Sigma_0$ than both $\math \Sigma_n$ and $\matb \Sigma$. However, for both shrinkage approaches using either $\mat I$ or $\matb \Sigma$, optimal tuning of $\alpha$ is application-specific and there is no one-size-fits-all strategy. 

For the general robustness principle \eqref{eq:rule-robust-estimation}, to find the minimal value of the local robustness measure $l$, the following min-max estimation framework can be formulated \cite{zhu2022generalized,liu2023robust}
\begin{equation}\label{eq:min-max-robust-estimation}
    \inf_{\vec \theta^*(\cdot)} \quad \sup_{
    \P_{\vec \xi} \in \cal B_{\epsilon}(\P_{0, \vec \xi}),~ \P_{0, \vec \xi} \in \cal F}
    h[\vec \theta^*(\P_{\vec \xi}), \vec \theta_0(\P_{0, \vec \xi})],
\end{equation}
where $\vec \theta_0(\P_{0, \vec \xi})$ denotes the true parameter of interest of the clean data distribution $\P_{0, \vec \xi}$, $\cal F$ is a possible distributional family where $\P_{0, \vec \xi}$ lies in, $\vec \theta^*(\P_{\vec \xi})$ the robust estimator of $\vec \theta_0(\P_{0, \vec \xi})$ based on the contaminated data-generating distribution $\P_{\vec \xi}$, and $h(\cdot, \cdot)$ a cost function (e.g., $\|\vec \theta^*(\P_{\vec \xi}) - \vec \theta_0(\P_{0, \vec \xi})\|$). The formulation \eqref{eq:min-max-robust-estimation} means that the clean data distribution is included in $\cal F$, and the data-generating distribution $\P_{\vec \xi}$ is perturbed from $\P_{0, \vec \xi}$. Since $\P_{0, \vec \xi}$ is unknown, the uniform upper bound (of the cost) over $\P_{0, \vec \xi}$ on $\cal F$ is studied, and the robust estimator $\vec \theta^*(\cdot)$ tries to reduce this uniform upper bound. For the partial contamination case under $d_{\text{TV}}$, see \cite{zhu2022generalized,gao2019robust}; for the full contamination case under $d_{\text{W}}$, see \cite{zhu2022generalized,liu2023robust}. Specifically, \cite{zhu2022generalized} provides a unified analytical framework of the robust estimator \eqref{eq:min-max-robust-estimation}, certifying that \eqref{eq:min-max-robust-estimation} can achieve the robustness property \eqref{eq:rule-robust-estimation}. To computationally efficiently solve \eqref{eq:min-max-robust-estimation} in some parametric cases (e.g., mean and covariance estimation for Gaussian distributions), \cite{gao2019robust} proposes to use the total-variation generative adversarial networks (TV-GANs) under $d_{\text{TV}}$, while \cite{liu2023robust} employs the Wasserstein GANs (W-GANs) under $d_{\text{W}}$. Let the distribution family under investigation is $\{\P_{\vec \xi}(\vec \theta)\}$, indexed by the parameter $\vec \theta$. 
Suppose the clean data distribution is $\P_{\vec \xi}(\vec \theta_0)$. Our aim is to robustly estimate the unknown parameter $\vec \theta_0$ using the empirical distribution $\Ph_{n, \vec \xi}$ of $\P_{\vec \xi}$. Under the projection-based estimation framework (i.e., minimum distance approach) \cite{donoho1988automatic,basu2011statistical,zhu2022generalized,liu2023robust}, the parameter estimation problem can be stated as
\[
\vec \theta_0 \defeq \argmin_{\vec \theta \in \R^m} d(\P_{\vec \xi}(\vec \theta), \P_{\vec \xi}),
\]
that is, projecting $\P_{\vec \xi}$ onto the parametric class $\{\P_{\vec \xi}(\vec \theta)\}_{\vec \theta \in \R^m}$, where $d$ denotes either the Kullback–Leibler divergence (under which the maximum likelihood estimation is recovered) \cite{basu2011statistical}, or the Wasserstein distance $d_{\text{W}}$ \cite{liu2023robust}, or the TV distance $d_{\text{TV}}$ \cite{gao2019robust}. Note that the data-generating distribution $\P_{\vec \xi}$ may not be included in $\{\P_{\vec \xi}(\vec \theta)\}_{\vec \theta \in \R^m}$, although $d(\P_{\vec \xi}, \P_{\vec \xi}(\vec \theta_0)) \le \epsilon$. In practice, we replace $\P_{\vec \xi}$ with its data-driven estimate $\Ph_{n, \vec \xi}$, yielding 
\[
\min_{\vec \theta \in \R^m} d(\P_{\vec \xi}(\vec \theta), \Ph_{n, \vec \xi}).
\]
The variational representation of the $1$-Wasserstein distance $d_{\text{W}}$ transforms the above display into
\begin{equation}\label{eq:W-GAN}
(\vec \theta^*, f^*) \defeq \argmin_{\vec \theta \in \R^m} \argmax_{f: \|f\|_{\text{Lip}} \le 1} \left[\E_{\Ph_{n, \vec \xi}} f(\vec \xi) - \E_{\P_{\vec \xi}(\vec \theta)} f(\vec \xi)\right],
\end{equation}
where $\|f\|_{\text{Lip}} \le 1$ means that $f$ is $1$-Lipschitz continuous. Formulation \eqref{eq:W-GAN} is called a W-GAN, in the sense that the generator mimics the data-generating distribution $\P_{\vec \xi}$ using $\P_{\vec \xi}(\vec \theta)$ by choosing the best parameter $\vec \theta^*$, while the discriminator differentiates $\P_{\vec \xi}(\vec \theta)$ from $\P_{\vec \xi}$ by choosing the best classifier $f^*$; note that in W-GANs, the function space $\{f: \|f\|_{\text{Lip}} \le 1\}$ is realized by a proper neural network. In \cite{liu2023robust}, it is shown that 
\begin{equation}\label{eq:W-robust-estimate}
\|\vec \theta^* - \vec \theta_0\| \le c\sqrt{\frac{m}{n}} + l \epsilon,
\end{equation}
in probability, for some finite constants $c$ and $l$, thus achieving the robustness property in \eqref{eq:rule-robust-estimation}. For technical details, see \cite{liu2023robust}. For the case for partial contamination under $d_{\text{TV}}$, see \cite{gao2019robust}, in which a similar robustness result in \eqref{eq:W-robust-estimate} is established, through employing another type of neural network.

\begin{figure}[!htbp]
    \centering
    \includegraphics[width=4.5cm]{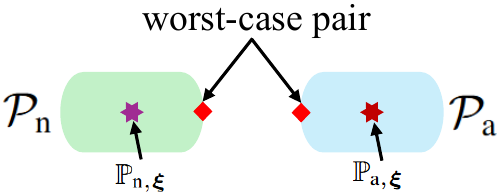}
    \caption{Illustration of the worst-case pair across two distributional uncertainty sets, consisting of the two distributions that are most difficult to distinguish.}
    \label{fig:worst-case-pair}
\end{figure}

Next, we introduce robust hypothesis testing under the Neyman-Pearson framework \cite{fauss2021minimax}, as an example, due to its wide applications in WSC and general engineering. For other robust testing frameworks, see, e.g., \cite{staudte1990robust,huber2009robust,basu2011statistical}. Let $\P_{\vec \xi}$ be the data-generating distribution, from which a sample $\vec \xi$ is drawn. Consider the classical testing problem 
\begin{equation}\label{eq:NP-test}
    \left\{
    \begin{array}{l}
         H_0: \P_{\vec \xi} = \P_{\text{n}, \vec \xi}, \\
         H_1: \P_{\vec \xi} = \P_{\text{a}, \vec \xi}. \\
    \end{array}
    \right.
\end{equation}
We aim to find a test function $\phi: \Xi \to [0, 1]$, where $\phi(\vec \xi)$ denotes the probability of rejecting $H_0$: that is, $\phi(\vec \xi) = 1$ means certainly accepting $H_1$, while $\phi(\vec \xi) = 0$ means certainly accepting $H_0$. Under this setting, the Type-I error is $\E_{\P_{\text{n}, \vec \xi}}\phi(\vec \xi)$, while the Type-II error is $\E_{\P_{\text{a}, \vec \xi}}[1-\phi(\vec \xi)]$. Consequently, the Neyman–Pearson testing can be formulated as the following optimization \cite{fauss2021minimax}
\begin{equation}\label{eq:NP-test-opt}
    \begin{array}{cl}
        \displaystyle \min_{\phi(\cdot)} & \E_{\P_{\text{a}, \vec \xi}} [1 -\phi(\vec \xi)] \\
         \st        & \E_{\P_{\text{n}, \vec \xi}}\phi(\vec \xi) \le \alpha,
    \end{array}
\end{equation}
where $\alpha \in [0, 1]$ denotes an error rate. The optimal test function is given as follows:
\begin{equation}\label{eq:NP-test-func}
    \phi(\vec \xi) = \left \{
     \begin{array}{cl}
         1, & \frac{p_{\text{a}}(\vec \xi)}{p_{\text{n}}(\vec \xi)} > k,\\
         \gamma, &  \frac{p_{\text{a}}(\vec \xi)}{p_{\text{n}}(\vec \xi)} = k,\\
         0, & \text{otherwise},
     \end{array}
    \right.
\end{equation}
where $\gamma \in [0, 1]$ is a random decision; $p_{\text{a}}(\vec \xi)$ and $p_{\text{n}}(\vec \xi)$ denote the probability density functions of $\P_{\text{a}, \vec \xi}$ and $\P_{\text{n}, \vec \xi}$, respectively; $k$ is such that $\E_{\P_{\text{n}, \vec \xi}}\phi(\vec \xi) = \alpha$. In practice, however, the probability models $\P_{\text{a}, \vec \xi}$ and $\P_{\text{n}, \vec \xi}$ can rarely be exactly specified. Hence, the testing problem \eqref{eq:NP-test} can be robustified as
\begin{equation}\label{eq:robust-NP-test}
    \left\{
    \begin{array}{l}
         H_0: \P_{\vec \xi} \in \cal P_{\text{n}}, \\
         H_1: \P_{\vec \xi} \in \cal P_{\text{a}}, \\
    \end{array}
    \right.
\end{equation}
for two uncertainty sets, for example, $\cal P_{\text{n}} \defeq \cal B_{\epsilon}(\P_{\text{n}, \vec \xi})$ and $\cal P_{\text{a}} \defeq \cal B_{\epsilon}(\P_{\text{a}, \vec \xi})$. As a result, considering the principle of the min-max distributional robustness defined before, \eqref{eq:NP-test-opt} can be modified into
\begin{equation}\label{eq:robust-NP-test-opt}
    \begin{array}{cl}
        \displaystyle \min_{\phi(\cdot)} \max_{\P \in \cal P_{\text{a}}} & \E_{\P} [1-\phi(\vec \xi)] \\
         \st        & \displaystyle \max_{\Q \in \cal P_{\text{n}}} \E_{\Q}\phi(\vec \xi) \le \alpha,
    \end{array}
\end{equation}
to obtain worst-case optimality and worst-case feasibility. Here, $\cal P_{\text{a}}$ and $\cal P_{\text{n}}$ can be defined using the Kullback–Leibler divergence, the Huber's contamination model \eqref{eq:huber-contamination-model}, the Wasserstein distance, etc \cite{fauss2021minimax}. The robust test function solving \eqref{eq:robust-NP-test-opt} is
\begin{equation}\label{eq:robust-NP-test-func}
    \phi^*(\vec \xi) = \left \{
     \begin{array}{cl}
         1, & \frac{p^*_{\text{a}}(\vec \xi)}{p^*_{\text{n}}(\vec \xi)} > k^*,\\
         \gamma^*, & \frac{p^*_{\text{a}}(\vec \xi)}{p^*_{\text{n}}(\vec \xi)} = k^*,\\
         0, & \text{otherwise},\\
     \end{array}
    \right.
\end{equation}
where $p^*_{\text{a}}$ and $p^*_{\text{n}}$ denote the worst-case distributions in $\cal P_{\text{a}}$ and $\cal P_{\text{n}}$, respectively, that are most difficult to distinguish from each other; for technical details and the determination method of $\gamma^* \in [0, 1]$, $k^*$, $p^*_{\text{a}}$ and $p^*_{\text{n}}$, see \cite{fauss2021minimax}. As indicated in \eqref{eq:robust-NP-test-func}, considering uncertainties in hypothesis testing essentially entails finding the worst-case distribution pair $(p^*_{\text{a}} \in \cal P_{\text{a}},~p^*_{\text{n}} \in \cal P_{\text{n}})$ that are most difficult to distinguish, i.e., the pair with the maximum similarity (minimum discrepancy). For extensive readings on robust hypothesis testing, refer to \cite{fauss2021minimax}. For a visual illustration, see Fig. \ref{fig:worst-case-pair}.

Last, we discuss robust regression techniques. Consider a data-generating process
\begin{equation}\label{eq:regression}
    \rvec s = \vec w_0 (\rvec x) + \rvec v_0,
\end{equation}
where $\rvec v_0$ denotes the clean noise distribution (e.g., pure Gaussian) with zero mean and covariance $\mat R_0$. Let $\P_{0, \rvec s, \rvec x}$ denote the clean-data joint distribution that defined by $\vec w_0$ and $\rvec v_0$ through \eqref{eq:regression}, and $\P_{\rvec s, \rvec x}$ the data-generating distribution that is contaminated from $\P_{0, \rvec s, \rvec x}$. For example, $\P_{\rvec s, \rvec x}$ can be generated by  $\vec w_0$ and an outlier-contaminated noise distribution $\rvec v$ through $ \rvec s = \vec w_0 (\rvec x) + \rvec v$. A regression problem leveraging $\P_{\rvec s, \rvec x}$ can be formulated as
\begin{equation}\label{eq:regression-M-esti}
    \vec w^* \defeq \argmin_{\vec w \in \cal W} \E_{\P_{\rvec s, \rvec x}} h(\rvec s, \vec w (\rvec x)),
\end{equation}
where $\cal W$ denotes a function space and $h$ a cost function. Robust regression has wide applications in WSC, including robust localization \cite[p.~67]{zoubir2012robust}, robust receiver design \cite{wang2025distributionallycombining}, to name a few. Note that \eqref{eq:regression-M-esti} is a special case of M-estimation \eqref{eq:M-estimation}. Therefore, all the previous robust methods based on M-estimation can be applied to conduct robust regression \eqref{eq:regression-M-esti}. Below, we consider a special case, i.e., the linear regression with $\mat R_0 \defeq \mat I$, to illustrate the robustification procedure. 


When the true data-generating distribution $\P_{\rvec s, \rvec x}$ is contaminated from the clean data distribtion $\P_{0, \rvec s, \rvec x}$ under $d_{\text{TV}}$, as in \eqref{eq:M-estimation-mean-cov}, we have the outlier-robust linear regression formulation \cite[Section~7.2]{huber2009robust}, \cite[p.~65]{zoubir2012robust}
\begin{equation}\label{eq:regression-M-esti-1}
    \min_{\mat W} \E_{\P_{\rvec s, \rvec x}} \rho(\rvec s - \mat W \rvec x),
\end{equation}
where the function $\rho$ is employed to combat outliers. In this case, as discussed before, the key is to ensure that the derivative $\nabla_{\mat W} \rho(\rvec s - \mat W \rvec x)$ is bounded. An alternative to \eqref{eq:regression-M-esti-1}, motivated by \cite{maronna1976robust}, is to study 
\begin{equation}\label{eq:regression-M-esti-2}
\min_{\mat W} \E_{\P_{\rvec s, \rvec x}} \rho([\rvec s - \mat W \rvec x]^\T [\rvec s - \mat W \rvec x]),
\end{equation}
for which case, $\nabla_{\mat W} \rho([\rvec s - \mat W \rvec x]^\T [\rvec s - \mat W \rvec x])$ needs to be bounded; cf. \eqref{eq:M-estimation-mean-cov}. It is straightforward to see that, for \eqref{eq:regression-M-esti-2}, $\rho$ should decay faster than $\rvec s - \mat W \rvec x$; otherwise, the derivative cannot be bounded when outliers occur (i.e., when the value of $\rvec s - \mat W \rvec x$ is large). In addition to M-estimation, other outlier-robust linear regression methods include \cite{diakonikolas2019efficient}, which is computationally efficient for high-dimensional problems.

If the contamination is under $d_{\text{W},p}$, as in \eqref{eq:wasserstein-mean-cov-estimation}, one robustification strategy is \cite{wang2025distributionallycombining}
\begin{equation}\label{eq:robust-linear-reg-Wasserstein-l2}
    \min_{\mat W} \max_{\P \in \cal B_{\text{W},p, \epsilon}(\P_{\rvec s, \rvec x})} \Tr \E_{\P} [\rvec s - \mat W \rvec x] [\rvec s - \mat W \rvec x]^\T.
\end{equation}
The quadratic function can be replaced with the absolute value function to simultaneously combat outliers, leading to \cite[Chapter~4]{chen2020distributionally}
\begin{equation}\label{eq:robust-linear-reg-Wasserstein-l1}
    \min_{\mat W} \max_{\P \in \cal B_{\text{W},p, \epsilon}(\P_{\rvec s, \rvec x})} \E_{\P} \vec 1^\T |\rvec s - \mat W \rvec x|,
\end{equation}
where $\vec 1$ denotes an all-one vector and $|\cdot|$ returns component-wise absolute values. In \cite{blanchet2019robust}, a more general strategy 
\begin{equation}
    \min_{\mat W} \max_{\P \in \cal B_{\text{W},p, \epsilon}(\P_{\rvec s, \rvec x})} \E_{\P} h(\mat W, \rvec s, \rvec x)
\end{equation}
is investigated, where $h$ can take diverse forms, such as the quadratic function and absolute value function. In addition, W-GANs can also be leveraged to robustly estimate $\mat W$ \cite{liu2023robust}:
\begin{equation}\label{eq:W-GAN-regression}
\mat W^* \defeq \argmin_{\mat W} \max_{f: \|f\|_{\text{Lip}} \le 1} \left[\E_{\P_{\rvec s, \rvec x}} f(\vec \xi) - \E_{\P_{\rvec s, \rvec x}(\mat W)} f(\vec \xi)\right],
\end{equation}
which brings computational efficiency for high-dimensional problems; cf. \eqref{eq:W-GAN}.

For both contamination cases, either under $d_{\text{TV}}$ or $d_{\text{W},p}$, the technical aim is to achieve $\|\mat W^* - \mat W_0\| \le l \epsilon$ for some finite constants $l$, i.e., satisfying the robustness property \eqref{eq:rule-robust-estimation} \cite{liu2023robust,diakonikolas2019efficient}. However, in favor of computational efficiency, a compromised error bound $e(\epsilon) \ge \epsilon$ may be unavoidable \cite{diakonikolas2019efficient}.

\subsubsection{Robust Optimization}
Optimization plays a central role in WSC (and general engineering) because many WSC problems can be formulated as decision-making models \cite{liu2024survey,wang2025distributionallybeamforming}. This importance also extends to statistics and its robust variants; recall \eqref{eq:M-estimation} and \eqref{eq:min-max-robust-estimation}. Depending on the decision vector $\vec x$ and the parameter vector $\vec \xi$, a decision-making model can be written as 
\begin{equation}\label{eq:opt}
    \begin{array}{cl}
       \displaystyle \min_{\vec x \in \cal X}  & h(\vec x, \vec \xi) \\
       \st  &  \vec g(\vec x, \vec \xi) \le \vec 0,
    \end{array}
\end{equation}
where $h$ defines the objective and $\vec g$ a collection of constraints. To be specific, for example, in resource allocation for wireless communications, $\vec x$ can denote a power allocation vector, while $\vec \xi$ the channel state information \cite{liu2024survey}. For another example, in adaptive beamforming for wireless sensing (e.g., estimating signal's direction of arrival), $\vec x$ can denote a receive beamformer, while $\vec \xi$ the steering vector \cite{wang2025distributionallybeamforming}. Note that $h$ and $\vec g$ can only rely on some components of $\vec \xi$. In practice, however, the parameter $\vec{\xi}$ is typically subject to uncertainty \cite{ben2009robust,vorobyov2013principles,wang2025uncertainty}. To be specific, in WSC, channel state information, sensor deployment locations, target azimuth and elevation angles, and available radio resources may not be precisely known \cite{wang2025uncertainty}. Ignoring these sources of uncertainty can significantly deteriorate the trustworthiness and efficiency of the resulting decisions \cite{ben2009robust,wang2025uncertainty}. Hence, robust decision-making under uncertainties must be studied. Two principal frameworks are robust optimization \cite{ben2009robust} and distributionally robust optimization (DRO) \cite{kuhn2025distributionally}. The former reached maturity in the 1990s, while the latter has advanced significantly since the 2000s. Below, we review the motivations, theories, and methods behind the two frameworks.

\begin{figure}[htbp]
    \centering
    \subfigure[Deterministic: Exact value known]{
        \begin{minipage}[htbp]{0.45\linewidth}
            \centering
            \includegraphics[height=1.5cm]{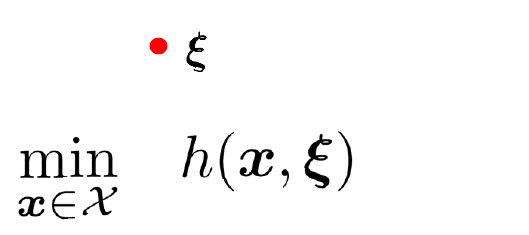}
        \end{minipage}
    }
    \subfigure[Robust: Domain known]{
        \begin{minipage}[htbp]{0.45\linewidth}
            \centering
            \includegraphics[height=1.5cm]{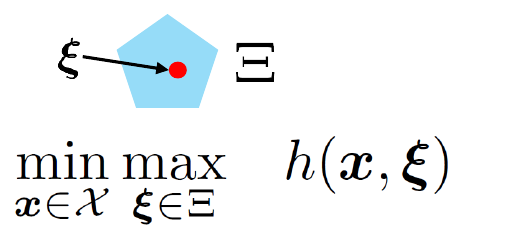}
        \end{minipage}
    }

    \subfigure[Stochastic: Distribution known]{
        \begin{minipage}[htbp]{0.45\linewidth}
            \centering
            \includegraphics[height=1.5cm]{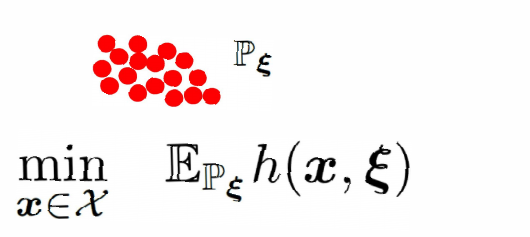}
        \end{minipage}
    }
    \subfigure[Distributionally Robust]{
        \begin{minipage}[htbp]{0.45\linewidth}
            \centering
            \includegraphics[height=1.5cm]{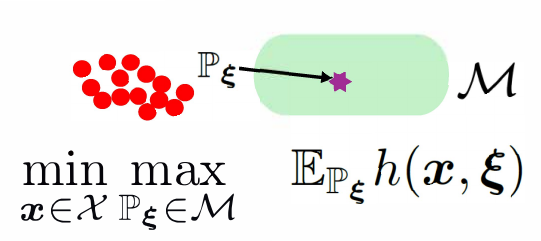}
        \end{minipage}
    }
    \caption{Visualized comparison among uncertainty-aware optimizations.}
    \label{fig:opt}
\end{figure}

Facing the uncertainty in $\vec \xi$, there exist two streams of technical treatments to robustify \eqref{eq:opt}, according to whether $\vec \xi$ is random. 
In the first stream, $\vec \xi$ is believed to be deterministic, although unknown to modelers. Suppose that the \textit{uncertainty set} of $\vec \xi$ is $\Xi$, usually a ball $\cal B_\epsilon(\vech \xi)$ centered at some points $\vech \xi$, the robustification of \eqref{eq:opt} can be formulated as \cite{ben2009robust}
\begin{equation}\label{eq:robust-opt}
    \begin{array}{cl}
       \displaystyle \min_{\vec x \in \cal X} \max_{\vec \xi \in \Xi}  & h(\vec x, \vec \xi) \\
       \st  & \displaystyle \max_{\vec \xi \in \Xi} \vec g(\vec x, \vec \xi) \le \vec 0,
    \end{array}
\end{equation}
which pursues worst-case optimality and feasibility. Conceptually, \eqref{eq:robust-opt} states that at the robust decision $\vec x^*$, the objective value $h(\vec x^*, \vec \xi)$ will not be worse than $\max_{\vec \xi \in \Xi} h(\vec x^*, \vec \xi)$ and the constraints will be uniformly satisfied [i.e., $\vec g(\vec x^*, \vec \xi) \le \vec 0$], for any realizations of $\vec \xi \in \Xi$. A key underlying assumption here is that the worst-case cost $\max_{\vec \xi \in \Xi} h(\vec x^*, \vec \xi)$ would be satisfactory in real-world operations; i.e., $h(\vec x^*, \vec \xi_0) \le \max_{\vec \xi \in \Xi} h(\vec x^*, \vec \xi) \le t$ for the true value $\vec \xi_0 \in \Xi$ and some prespecified performance thresholds $t$. For more motivations and clarifications of \eqref{eq:robust-opt}, see \cite{ben2009robust}. In addition to the above conventional interpretations, another new understanding can be coherently drawn from the proposed robustness theory; recall \eqref{eq:decision-making-robustness-min-max-true} and \eqref{eq:decision-making-robustness-min-max}, as well as their contextual interpretations. To be short, the aim is to find the robust decision $\vec x^*$ and the one-sided robustness measure $l^*$ such that
\[
    h(\vec x^*, \vec \xi) \le h(\vec x_0, \vec \xi_0) + l^* \epsilon,~~~\forall \vec \xi \in \cal B_\epsilon(\vec \xi_0);
\]
see \eqref{eq:decision-making-robustness-one-sided}. In the second stream, it is acknowledged that $\vec \xi$ is naturally a random vector and obeys the probability distribution $\P_{\vec \xi}$. Consequently, stochastic programming
\begin{equation}\label{eq:sto-program}
    \begin{array}{cl}
       \displaystyle \min_{\vec x \in \cal X}  & \E_{\P_{\vec \xi}} h(\vec x, \vec \xi) \\
       \st  & \E_{\P_{\vec \xi}} \vec g(\vec x, \vec \xi) \le \vec 0,
    \end{array}
\end{equation}
and chance-constrained optimization
\begin{equation}\label{eq:chance-constraints}
    \begin{array}{cl}
       \displaystyle \min_{\vec x \in \cal X}  & \E_{\P_{\vec \xi}} h(\vec x, \vec \xi) \\
       \st  & \operatorname{Pr}_{\P_{\vec \xi}} [\vec g(\vec x, \vec \xi) \le \vec 0] \ge \alpha,
    \end{array}
\end{equation}
where $\alpha \in [0, 1]$ is a chance-constraint level (e.g., $0.95$), serve as standard variants \cite{shapiro2021lectures}. In \eqref{eq:sto-program} and \eqref{eq:chance-constraints}, the objective functions are specified by the mean of $h(\vec x, \vec \xi)$. However, they can also be defined using \textit{risk measures} of $h(\vec x, \vec \xi)$ \cite[Section~5]{kuhn2025distributionally}. For example, the objective function can be any $\beta$-quantile of $h(\vec x, \vec \xi)$, for $\beta \in (0, 1)$. As a result, the corresponding objective function is called $\beta$-Value-at-Risk ($\beta$-VaR). For more possibilities of objective functions, see \cite[Table~1]{wang2025uncertainty}, \cite[Section~5]{kuhn2025distributionally}. When the distribution $\P_{\vec \xi}$ is defined by a physical process, law, or model, \eqref{eq:sto-program} and \eqref{eq:chance-constraints} are called \textit{model-driven}. In contrast, if $\P_{\vec \xi}$ is approximated by an empirical distribution constructed using historical observations, they are termed \textit{data-driven}.
In practice, accurately specifying the true distribution $\P_{0, \vec \xi}$ of $\vec \xi$ is difficult. Hence, the min-max robustification can also be conducted over $\P_{\vec \xi}$, leading \eqref{eq:sto-program} to its distributionally robust counterpart \cite{kuhn2025distributionally}
\begin{equation}\label{eq:DRO}
    \begin{array}{cl}
       \displaystyle \min_{\vec x \in \cal X} \max_{\P_{\vec \xi} \in \cal M}  & \E_{\P_{\vec \xi}} h(\vec x, \vec \xi) \\
       \st  & \displaystyle \max_{\P_{\vec \xi} \in \cal M} \E_{\P_{\vec \xi}} \vec g(\vec x, \vec \xi) \le \vec 0.
    \end{array}
\end{equation}
Here, although we do not exactly know the true distribution $\P_{0, \vec \xi}$, we assumed that $\P_{0, \vec \xi}$ is included in the uncertainty set $\cal M$. As for \eqref{eq:chance-constraints}, the constraint can be robustified as
$
\min_{\P_{\vec \xi} \in \cal M} \operatorname{Pr}_{\P_{\vec \xi}} [\vec g(\vec x, \vec \xi) \le \vec 0] \ge \alpha
$. 
In addition to the interpretation of \eqref{eq:DRO} based on worst-case optimality and feasibility \cite{kuhn2025distributionally}, another new understanding can be coherently drawn from the proposed robustness theory; recall \eqref{eq:decision-making-dist-robustness-min-max-true} and \eqref{eq:decision-making-dist-robustness-min-max}, as well as their contextual interpretations. The aim is to find the robust decision $\vec x^*$ and the one-sided robustness measure $l^*$ such that
\[
    \E_{\P_{\vec \xi}} h(\vec x^*, \vec \xi) \le \E_{\P_{0, \vec \xi}} h(\vec x_0, \vec \xi) + l^* \epsilon,~~~\forall \P_{\vec \xi} \in \cal B_\epsilon(\P_{0, \vec \xi});
\]
see \eqref{eq:dist-local-decision-making-robustness-one-sided}. For a visual summary of uncertainty-aware optimizations, see Fig. \ref{fig:opt}.

Note that the traditional robust optimization \eqref{eq:robust-opt} is mathematically a special case of the DRO \eqref{eq:DRO} if we restrict $\P_{\vec \xi}$ to point-mass distributions. Therefore, in what follows, we focus on technical discussions of \eqref{eq:DRO}.

Although philosophically attractive, the DRO model \eqref{eq:DRO} has its inherent operational intricacies in practice \cite{kuhn2025distributionally}:
\begin{itemize}
    \item The distributional uncertainty set $\mathcal{M}$ plays a crucial role in characterizing uncertainties, yet its design is inherently problem-specific and lacks universal principles. Different choices of uncertainty sets can lead to substantially different performance across real-world applications. Recall from robust statistics that to combat data outliers, uncertainty sets based on $d_{\text{TV}}$ should be used, while to fight against full data contamination, those leveraging $d_{\text{W},p}$ are favored. In the context of DRO, typical choices of $\cal M$ include moment-based sets, $\phi$-divergence (or called $f$-divergence) balls, TV balls, Wasserstein balls, etc.

    \item The problem \eqref{eq:DRO}, in its general form, is computationally challenging because the inner maximization is an infinite-dimensional optimization. The complexity further escalates when the decision variable $\vec x$ and the data variable $\vec \xi$ are high-dimensional. Consequently, DRO admits efficient or analytic solutions only in specific cases. This motivates substantial effort toward developing computationally tractable reformulations or approximations.

    \item In addition to the computational difficulty, in the data-driven case where the uncertainty set $\cal M$ is constructed as a ball $\cal B_{\epsilon}(\Ph_{n, \vec \xi})$ centered at an $n$-sample empirical distribution $\Ph_{n, \vec \xi}$, exponentially many samples, in the dimension of $\vec \xi$, should be typically expected to ensure statistical efficiency.
\end{itemize}

One important property of \eqref{eq:DRO} is its equivalence to regularization effects, thus linking itself to trustworthy machine learning techniques. To be specific, supposing that $\cal M$ is constructed using a ball $\cal B_{\epsilon}(\Pb_{\vec \xi})$ centered at an estimate $\Pb_{\vec \xi}$ of the true distribution $\P_{0, \vec \xi}$, under some mild technical conditions, the DRO model \eqref{eq:DRO} is equivalent to a regularized formulation
\begin{equation}\label{eq:regularied}
    \begin{array}{cl}
       \displaystyle \min_{\vec x \in \cal X} & \E_{\Pb_{\vec \xi}} h(\vec x, \vec \xi)  + \epsilon f_1(\vec x) \\
       \st  & \displaystyle \E_{\Pb_{\vec \xi}} \vec g(\vec x, \vec \xi) + \epsilon \vec f_2(\vec x) \le \vec 0,
    \end{array}
\end{equation}
for some regularizers $f_1$ and $\vec f_2$; for some concrete examples, see \cite{kuhn2025distributionally,wang2025distributionallycombining,wang2025distributionallybeamforming}. In some cases (e.g., under weaker technical conditions), the equivalence between \eqref{eq:DRO} and \eqref{eq:regularied} cannot be guaranteed. However, the following inequality can still be established to build a (tight) upper bound of the true but unknown cost:
\[
\E_{\P_{0, \vec \xi}} h(\vec x, \vec \xi) \le \max_{\P_{\vec \xi} \in \cal B_{\epsilon}(\Pb_{\vec \xi})} \E_{\P_{\vec \xi}} h(\vec x, \vec \xi) \le \E_{\Pb_{\vec \xi}} h(\vec x, \vec \xi)  + \epsilon f_1(\vec x).
\]
As a result, \eqref{eq:regularied} serves as a relaxation of, and thus approximately solves, \eqref{eq:DRO} in the sense that the worst-case cost is determined by \eqref{eq:regularied} instead of \eqref{eq:DRO}. Another insight from \eqref{eq:regularied} is that $f_1$ and $\vec f_2$ can be interpreted as decision margins that buffer the distributional uncertainty in $\Pb_{\vec \xi}$ relative to $\P_{0, \vec \xi}$.

In data-driven settings where $\Pb_{\vec \xi} \defeq \Ph_{n, \vec \xi}$, \eqref{eq:regularied} particularizes into the empirical risk minimization (ERM) 
\begin{equation}\label{eq:regularied-SAA}
    \begin{array}{cl}
       \displaystyle \min_{\vec x \in \cal X} & \displaystyle \frac{1}{n} \sum^n_{i = 1} h(\vec x, \vec \xi_i)  + \epsilon f_1(\vec x) \\
       \st  & \displaystyle \frac{1}{n} \sum^n_{i = 1}  \vec g(\vec x, \vec \xi_i) + \epsilon \vec f_2(\vec x) \le \vec 0,
    \end{array}
\end{equation}
which builds the foundations of regularization-based robust machine learning techniques, including robust regression, classification, and deep learning \cite{shafieezadeh2017Regularization}.

Another important property of the DRO formulation \eqref{eq:DRO} is its inherent connections to adversarial learning in data-driven settings \cite{wang2025learning,bai2021recent,kuhn2025distributionally}. To be specific, under the $\phi$-divergence ball $\cal B_{\phi, \epsilon}(\Ph_{n, \vec \xi})$, \eqref{eq:DRO} is equivalent to \cite[Theorem~4]{wang2025learning}
\begin{equation}\label{eq:adversarial-reweighting}
    \begin{array}{cl}
       \displaystyle \min_{\vec x \in \cal X} \max_{\vec \mu \ge \vec 0}  & \displaystyle \sum^n_{i = 1} \mu_i h(\vec x, \vec \xi_i) \\
       \st  & \displaystyle \max_{\vec \mu \ge \vec 0} \sum^n_{i = 1} \mu_i \vec g(\vec x, \vec \xi_i) \le \vec 0 \\
       & d_{\phi}(\vec \mu \| \vech \mu) \le \epsilon, \\
    \end{array}
\end{equation}
where $d_{\phi}$ denotes the $\phi$-divergence of $\vec \mu$ from $\vech \mu$, $\vec \mu$ an adversarial data weight vector, and $\vech \mu \defeq [1/n, 1/n, \ldots, 1/n]$ the empirical data weight vector. Formulation \eqref{eq:adversarial-reweighting} is called \textit{adversarial reweighting} or \textit{hard sample mining} in machine learning because the empirical weight $1/n$ of the sample $\vec \xi_i$ is modified to $\mu_i$. In contrast, under the $p$-Wasserstein ball and some technical conditions, \eqref{eq:DRO} is equivalent to \cite[Theorem~5]{wang2025learning}, \cite[Section~3.3]{chen2020distributionally}, \cite{shafieezadeh2017Regularization}
\begin{equation}\label{eq:adversarial-learning-group}
    \begin{array}{cl}
       \displaystyle \min_{\vec x \in \cal X} \max_{\{\vec \zeta_i\}_i \subset \Xi}  & \displaystyle \frac{1}{n} \sum^n_{i = 1} h(\vec x, \vec \zeta_i) \\
       \st  & \displaystyle \max_{\{\vec \zeta_i\}_i \subset \Xi} \frac{1}{n} \sum^n_{i = 1} \vec g(\vec x, \vec \xi_i) \le \vec 0 \\
       & \displaystyle \frac{1}{n} \sum^n_{i = 1} c^p(\vec \zeta_i, \vec \xi_i) \le \epsilon^p, \\
    \end{array}
\end{equation}
where $c$ is the distance in defining the $p$-Wasserstein distance \eqref{eq:W-distance}. Under the $p$-Wasserstein ball and some other technical conditions, \eqref{eq:DRO} is equivalent to \cite[Theorem~4.18]{kuhn2025distributionally}
\begin{equation}\label{eq:adversarial-learning}
    \begin{array}{cl}
       \displaystyle \min_{\vec x \in \cal X} & \displaystyle \frac{1}{n} \sum^n_{i = 1} \max_{\vec \zeta_i \in \Xi:c(\vec \zeta_i, \vec \xi_i) \le \epsilon} h(\vec x, \vec \zeta_i) \\
       \st  & \displaystyle \frac{1}{n} \sum^n_{i = 1} \max_{\vec \zeta_i \in \Xi:c(\vec \zeta_i, \vec \xi_i) \le \epsilon} \vec g(\vec x, \vec \zeta_i) \le \vec 0.
    \end{array}
\end{equation}
Letting $\{\vec \zeta^*_i\}_{i}$ solve \eqref{eq:adversarial-learning-group} or \eqref{eq:adversarial-learning} with respect to the objectives, $\{\vec \zeta^*_i\}_{i}$ are called \textit{adversarial samples}. Both \eqref{eq:adversarial-learning-group} and \eqref{eq:adversarial-learning} are called \textit{adversarial learning}. The technical difference lies in the order of the summation and the maximization. In the practice of machine learning, particularly when implementing mini-batch stochastic gradient-based methods for minimization and maximization, the individual maximization formulation \eqref{eq:adversarial-learning} is preferred over the group maximization formulation \eqref{eq:adversarial-learning-group} because of its computational simplicity; see \cite{bai2021recent}.

Adversarial learning is more often than not criticized for its extreme conservatism, which impedes real-world operational performance. As such, mitigating strategies, such as \textit{data augmentation}, are proposed to balance between distributional information in training data and distributional robustness against dataset imperfection. One representative example is to modify the adversarial-learning objective as follows \cite{wang2025learning}:
\begin{equation}\label{eq:BDR}
\displaystyle \min_{\vec x \in \cal X} \displaystyle \beta \frac{1}{n} \sum^n_{i = 1} h(\vec x, \vec \xi_i) + \displaystyle (1-\beta) \frac{1}{n} \sum^n_{i = 1} \max_{\vec \zeta_i \in \Xi:c(\vec \zeta_i, \vec \xi_i) \le \epsilon} h(\vec x, \vec \zeta_i), 
\end{equation}
for a tradeoff coefficient $\beta \in [0, 1]$. In the above learning framework, $2n$ samples, including $n$ original samples $\{\vec \xi_i\}_{i}$ and $n$ adversarial samples $\{\vec \zeta^*_i\}_{i}$, are involved, from which data augmentation comes.

After DRO modeling and reformulation, the resulting regularized or adversarial learning counterparts are solved by off-the-shelf solvers or gradient-based methods, which, however, require additional algorithmic and computational endeavors \cite[Section~9]{kuhn2025distributionally}, \cite[Algorithm~1]{wang2025learning}, \cite[Eq.~(4)]{bai2021recent}.

\subsubsection{Robust Machine Learning}
Machine learning is concerned with studying the patterns in, drawing conclusions from, and making predictions or decisions based on data. It is largely grounded in statistics and optimization techniques. Statistics differs from machine learning in its technical focus, although both can tackle a shared set of real-world data-driven problems: the former highlights modeling and interpretation of data (i.e., understanding the characteristics of data sources), while the latter emphasizes making accurate predictions or decisions based on data (i.e., optimizing performance on downstream tasks). To be specific, for example, given the same medical dataset, a statistician investigates whether smoking causes heart disease, whereas a machine learning practitioner detects heart disease from patient records without analyzing causal relationships. Note that in the latter case, although smoking is acknowledged to be strongly correlated with heart disease, the causal relationship is disregarded (e.g., whether heart disease may cause smoking, or converse). For another example, in wireless communications, given pilot data, statistics examines whether the data conform to the assumed physical channel model (i.e., model validation and estimation), whereas machine learning aims to establish decision boundaries to classify transmitted symbols (i.e., detecting non-pilot payloads). In short, statistics explains what and why happened, while machine learning predicts what will happen. However, the boundary between statistics and machine learning is increasingly blurred in modern practice by, e.g., non-parametric methods in statistics and generative models in machine learning. Both statistics and machine learning problems are typically formulated into a data-driven optimization model to draw conclusions, generate predictions, and make decisions. As such, the robust statistics and optimization frameworks discussed above are naturally suited to enabling robust machine learning. To be specific, for example, regularized and adversarial learning-based robust regression, classification, deep learning, etc., against Wasserstein perturbations, can be readily obtained; see, e.g., \cite{chen2020distributionally,shafieezadeh2017Regularization}. For another example, machine learning methods that rely on outlier-robust mean and covariance estimation, e.g., Gaussian mixture models clustering, against TV perturbations, can be straightforwardly realized \cite{lai2016agnostic,diakonikolas2019robust}.

Typically, robust machine learning, as robust statistics does, aims to combat data uncertainties, or more rigorously speaking, distributional shifts, in prediction or decision-making. The primary technical focus is to achieve robustness, that is, the insensitivity or stability, of the risk $\E_{\P_{\vec \xi}} h(\vec x^*, \vec \xi)$ at the robust decision $\vec x^*$, against uncertainties in $\P_{\vec \xi}$. In machine learning, two concepts, \textit{generalizability} and robustness, need to be distinguished. Generalizability, which is one constituting aspect of robustness, specifically refers to in-distribution performance stability, from the finite-sample empirical risk $\E_{\Ph_{n, \vec \xi}} h(\vec x^*, \vec \xi)$ to the infinite-sample population risk $\E_{\P_{\vec \xi}} h(\vec x^*, \vec \xi)$. In contrast, robustness also extends to out-of-distribution performance stability, from the contaminated-population risk $\E_{\P_{\vec \xi}} h(\vec x^*, \vec \xi)$ to the clean-population risk $\E_{\P_{0, \vec \xi}} h(\vec x^*, \vec \xi)$. Hence, a machine learning scheme is robust if it can generate a decision $\vec x^*$ such that both the in-distribution generalization error
\begin{equation}
\E_{\P_{\vec \xi}} h(\vec x^*, \vec \xi) - \E_{\Ph_{n, \vec \xi}} h(\vec x^*, \vec \xi)
\end{equation}
and the out-of-distribution extension error
\begin{equation}
\E_{\P_{0, \vec \xi}} h(\vec x^*, \vec \xi) - \E_{\P_{\vec \xi}} h(\vec x^*, \vec \xi)
\end{equation}
are simultaneously small; compare with the conceptual counterparts in robust statistics in \eqref{eq:rule-robust-estimation}. Note that, in the deployment phase of a machine learning model, the relevant real-world testing distribution $\P_{0, \vec \xi}$ may deviate from the laboratory testing distribution $\P_{\vec \xi}$ used in the development phase. In addition, due to data scarcity, the empirical distribution $\Ph_{n, \vec \xi}$ in the training phase usually differs from its corresponding population distribution $\P_{\vec \xi}$ in the laboratory testing phase. Excerpt for robustness, the solution optimality is of natural importance as well, which is measured by the \textit{excess risk}
\begin{equation}
    \E_{\P_{0, \vec \xi}} h(\vec x^*, \vec \xi) - \E_{\P_{0, \vec \xi}} h(\vec x_0, \vec \xi).
\end{equation}
Therefore, a machine learning task seeks to reduce the in-distribution generalization error, the out-of-distribution extension error, and the excess risk, which can be realized by min-max distributionally robust optimization; recall \eqref{eq:dist-local-decision-making-robustness-one-sided} and \eqref{eq:decision-making-dist-robustness-min-max-true}.

Two kinds of distributional shifts of $\P_{\vec \xi}$ from $\P_{0, \vec \xi}$ are typical: non-adversarial data shifts and adversarial data shifts \cite{braiek2025machine}. The former refers to the natural and unintentional changes in the data-generating process, such as environmental variations and sensor biases, while the latter arises from intentionally and maliciously crafted perturbations that can mislead and degrade system performance, such as attacks and deliberately faked samples. As indicated before, these data shifts can be quantified using $d_{\text{TV}}$ distributional balls, $d_{\text{W}, p}$ distributional balls, or $d_{\phi}$ distributional balls, depending on whether the shifts are driven by partial-contamination outliers, full-contamination changes, or data-frequency (i.e., data-weight) manipulations, respectively. Below, we discuss some trending mitigating strategies in robust machine learning against distributional uncertainties, by remarking connections with robust statistics and distributionally robust optimization.

\textit{Regularized Learning}: Robust machine learning is concerned with trustworthy decision-making when the data-driven empirical distribution $\Ph_{n, \vec \xi}$ is employed as a surrogate for the unknown distribution $\P_{0, \vec \xi}$. To minimize $\E_{\P_{0, \vec \xi}} h(\vec x, \vec \xi)$, one possibility is to construct and minimize an upper bound of the form $\E_{\Ph_{n, \vec \xi}} h(\vec x, \vec \xi) + \lambda f(\vec x)$, where $f(\vec x)$ is a regularizer. Note that the empirical estimate $\E_{\Ph_{n, \vec \xi}} h(\vec x, \vec \xi)$ cannot serve as an upper bound for $\E_{\P_{0, \vec \xi}} h(\vec x, \vec \xi)$, because the expectation of the random quantity $\E_{\Ph_{n, \vec \xi}} h(\vec x, \vec \xi)$ is exactly $\E_{\P_{0, \vec \xi}} h(\vec x, \vec \xi)$. To specify $f(\vec x)$, measure concentration inequalities, such as Hoeffding’s and Bernstein’s inequalities, among many others, can be leveraged \cite[Chapters~2–3]{wainwright2019high}. Alternatively, distributionally robust optimization can be employed \cite{shafieezadeh2017Regularization}; see \eqref{eq:regularied}. Under this principle of upper-bound exploitation, $f(\vec x)$ typically depends on properties of the cost function $h(\vec x, \vec \xi)$, such as boundedness or its Lipschitz norm. Another possibility is to incorporate some prior belief into the learning process. For example, one can control the dispersion of the random quantity $h(\vec x, \vec \xi)$ by letting $f(\vec x) \defeq \D_{\Ph_{n, \vec \xi}} h(\vec x, \vec \xi)$ where $\D$ denotes the variance operator \cite[Table~1]{wang2025uncertainty}. For another example, one can limit the complexity of the hypothesis space and thereby reduce overfitting on the training data. To specify $f(\vec x)$ in this setting, richness measures of hypothesis classes, such as the Vapnik–Chervonenkis (VC) dimension and Rademacher complexity, are studied \cite[Chapter~4]{wainwright2019high}. Specific examples under this principle of prior-belief embedding include Lasso (Least Absolute Shrinkage and Selection Operator) regression, Ridge regression (i.e., Tikhonov regularization), and weight decay in deep learning (i.e., norm regularization of weights).

\textit{Data Augmentation}: Data augmentation is another popular approach to improve the robustness of a machine learning model. Two fundamental philosophies lie behind data augmentation: a) one is to enrich the training dataset by augmenting new and possibly valid samples (i.e., they are possibly drawn from the population distribution, although unseen in the current training dataset); b) the other is to add adversarial samples into the training dataset so that the model can be robust against distributional shift. The former notion can be straightforwardly imagined in practice, while the latter notion can be rigorously justified using the learning scheme \eqref{eq:BDR}. Operationally, for both cases, data augmentation begins with manipulating existing training samples. To be specific, in image classification, one can conduct geometric transformations, such as random rotation, translation, and mirroring. In text recognition, one can apply synonym replacements, typo injection, rephrasing, and main-subordinate clauses swapping. In adversarial training, one can draw an adversarial sample from a norm ball of a training sample; see the adversarial maximization in \eqref{eq:BDR}. Modern generative models, such as generative adversarial nets (GANs) \cite{goodfellow2014generative} and generative pre-trained transformers (GPTs), are potential choices for advanced data augmentation.

\textit{Data Cleaning}: In addition to using data augmentation to enhance data diversity, another key practice in robust learning is purifying a contaminated dataset to ensure high data quality \cite{ilyas2019data}. Out-of-distribution detection and removal are related, but performed at testing time, whereas data cleaning occurs during training. This motivation parallels the framework of outlier-robust estimation under $d_{\text{TV}}$-partial contamination, where outliers are to be removed or damped \cite{diakonikolas2019robust,lai2016agnostic}, e.g., by using a redescending or monotonic influence function in M-estimation, respectively \cite{huber2009robust}; recall \eqref{eq:M-estimation-mean-cov}, \eqref{eq:weighted-robust-mean}, and \eqref{eq:weighted-robust-cov}.

\begin{figure}[htbp]
    \centering
    \subfigure[Redescending influence function]{
        \begin{minipage}[htbp]{0.45\linewidth}
            \centering
            \includegraphics[height=2.8cm]{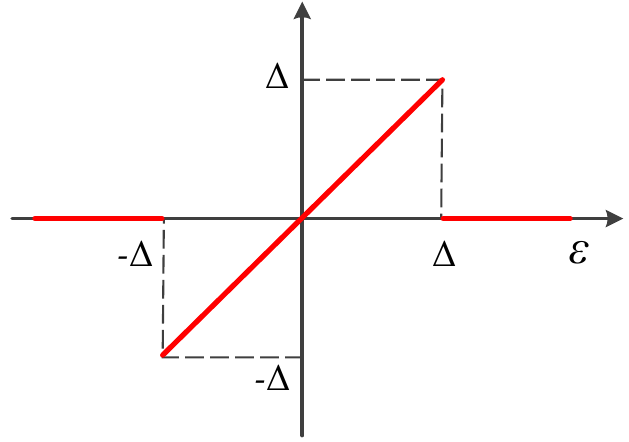}
        \end{minipage}
    }
    \subfigure[Monotonic influence function]{
        \begin{minipage}[htbp]{0.45\linewidth}
            \centering
            \includegraphics[height=2.8cm]{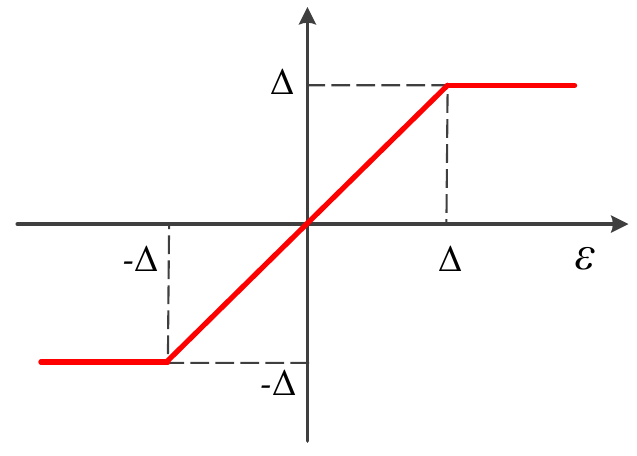}
        \end{minipage}
    }
    \caption{The influence functions used to identify, and remove or attenuate the outliers; $\epsilon$ denotes the difference between a sample and the clean-data mean; $\Delta$ indicates the censoring threshold. The larger the value of $\epsilon$, the more probable that the associated sample is an outlier. In (a), an outlier (having a value larger than $\Delta$) is totally removed. In (b), an outlier is attenuated to $\Delta$.}
    \label{fig:re-descending}
\end{figure}

\textit{Adversarial Learning}: Adversarial learning aims to achieve robustness of a machine learning model by employing adversarial samples \cite{goodfellow2015explaining}. Operationally, adversarial samples are constructed using norm balls centered at the available training data; see \eqref{eq:adversarial-learning-group}, \eqref{eq:adversarial-learning}, and \eqref{eq:BDR}, where $c(\vec \zeta, \vec \xi)$ is instantiated by a norm $\| \vec \zeta - \vec \xi \|$, such as the vector sup-norm or $2$-norm \cite{bai2021recent,wang2025learning}. The underlying assumption is that, in the real-world testing phase (i.e., the deployment stage), the operating distribution will be no worse than the adversarial distribution (generated by adversarial samples) constructed during the laboratory development phase. Although this assumption is not necessarily valid in practice, adversarial learning has nevertheless demonstrated satisfactory robustness across a wide range of real-world applications. Adversarial training \eqref{eq:adversarial-learning} can be implemented through gradient ascent and descent, alternately over $\{\vec \zeta_i\}_i$ and over $\vec x$, respectively, although the theoretically global optimality cannot be guaranteed. Yet, empirically, this strategy can result in significant robustness in machine learning \cite{goodfellow2015explaining,bai2021recent}. One representative instance of adversarial attacks, i.e., gradient ascent over $\{\vec \zeta_i\}_i$, is called \textit{fast gradient sign method}. To be specific, in \eqref{eq:adversarial-learning} and \eqref{eq:BDR}, for every $i$, the inner maximization is conducted through one-step (thus sub-optimal) gradient ascent \cite{goodfellow2015explaining}:
\begin{equation}\label{eq:fast-gradient-sign}
    \vec \zeta_i \leftarrow \vec \xi_i + \epsilon \cdot \operatorname{sign}(\nabla_{\vec \xi} h(\vec x, \vec \xi_i)),
\end{equation}
where $\operatorname{sign}(\cdot)$ returns $+1$ if the argument is positive and $-1$ if it is negative. A more general approach is based on projected gradient ascent: i.e., in \eqref{eq:adversarial-learning} and \eqref{eq:BDR}, for every $i$,
\begin{equation}\label{eq:projected-ascent-attack}
\left\{
\begin{array}{l}
     \vec \xi^+_i \leftarrow \vec \xi_i + \alpha \cdot \nabla_{\vec \xi} h(\vec x, \vec \xi_i), \\
     \vec \zeta_i \leftarrow \displaystyle \operatorname{proj}_{\{\vec \zeta_i \in \Xi:c(\vec \zeta_i, \vec \xi_i) \le \epsilon\}} (\vec \xi^+_i),
\end{array}
\right.
\end{equation}
where $\alpha \ge 0$ denotes a gradient-ascent rate and $\operatorname{proj}_{\cal A} (\vec a)$ means projecting the quantity $\vec a$ onto the space $\cal A$. The ascent operation in \eqref{eq:projected-ascent-attack} can be repeated multiple steps, as opposed to the one-step simplification in \eqref{eq:fast-gradient-sign}, to increase the degree of adversarial attacks. For a specific examination of adversarial learning in WSC, see \cite{adesina2022adversarial}.

\textit{Generative Adversarial Learning}: Generative adversarial learning aims to learn a crafted distribution $\Q_{\vec \xi}$ that mimics the underlying true data-generating distribution $\P_{\vec \xi}$, thereby improving the data richness, as data augmentation does. Generative adversarial learning can be seen as a game between a generator and a discriminator, where the generator tries to reduce the discrepancy between $\Q_{\vec \xi}$ and $\P_{\vec \xi}$, while the discriminator tries to differentiate samples from $\Q_{\vec \xi}$ versus $\P_{\vec \xi}$. The original version of generative adversarial learning, due to \cite{goodfellow2014generative}, is formulated as
\begin{equation}
    \min_{\vec G} \max_D \E_{\vec \xi \sim \P_{\vec \xi}}[\log D(\vec \xi)]+\E_{\vec z \sim \P_{\vec z}}[\log (1-D(\vec G(\vec z)))],
\end{equation}
or equivalently,
\begin{equation}
    \min_{\Q_{\vec \xi}} \max_D \E_{\vec \xi \sim \P_{\vec \xi}}[\log D(\vec \xi)]+\E_{\vec \xi \sim \Q_{\vec \xi}}[\log (1-D(\vec \xi))],
\end{equation}
or further equivalently,
\begin{equation}\label{eq:JS-GAN}
    \min_{\Q_{\vec \xi}} d_{\text{JS}}(\P_{\vec \xi} \| \Q_{\vec \xi}),
\end{equation}
where $\vec G$ is a generator that produces the new target distribution $\Q_{\vec \xi}$ from a source distribution $\P_{\vec z}$ under the data-generating law $\vec \xi = \vec G(\vec z)$, $D$ is a discriminator (i.e., a two-category classifier) between $\Q_{\vec \xi}$ and $\P_{\vec \xi}$, and $d_{\text{JS}}$ denotes the Jensen–Shannon divergence (a similarity measure of two distributions). In practice, both $\vec G$ and $D$ can be appropriate neural networks, so it gives generative adversarial nets (GANs). Later extensions to JS-GAN \eqref{eq:JS-GAN} include $\phi$-GAN (also known as $f$-GAN), TV-GAN, and Wasserstein-GAN, etc., where $d_{\text{JS}}$ is replaced with $d_{\phi}$, $d_{\text{TV}}$, and $d_{\text{W}}$, respectively; see \cite{gao2019robust,liu2023robust}. Taking TV-GAN as an example, we have
\begin{equation}
    \min_{\Q_{\vec \xi}} d_{\text{TV}}(\P_{\vec \xi}, \Q_{\vec \xi}),
\end{equation}
or equivalently, 
\begin{equation}\label{eq:TV-GAN}
\min_{\Q_{\vec \xi}} \max_{f: \|f\|_{\infty} \le 1} \left[\E_{\P_{\vec \xi}} f(\vec \xi) - \E_{\Q_{\vec \xi}} f(\vec \xi)\right]
\end{equation}
under the variational representation of the TV distance; $\|f\|_{\infty} \le 1$ means that the sup-norm of the function $f$ is one, i.e., $\sup_{\vec \xi \in \Xi} |f(\vec \xi)| \le 1$. Here, $f$ acts as a discriminator to differentiate between $\P_{\vec \xi}$ and $\Q_{\vec \xi}$; NB: the best discriminator can maximize the difference between $\E_{\P_{\vec \xi}} f(\vec \xi)$ and $\E_{\Q_{\vec \xi}} f(\vec \xi)$. In real-world operation, $\P_{\vec \xi}$ is approximated by its data-driven empirical estimate $\Ph_{n, \vec \xi}$ and $\Q_{\vec \xi}$ is represented by a parametric family $\P_{\vec \xi}(\vec \theta)$, leading \eqref{eq:TV-GAN} to \cite{gao2019robust}
\begin{equation}\label{eq:TV-GAN-practice}
\min_{\vec \theta \in \R^m} \max_{f: \|f\|_{\infty} \le 1} \left[\E_{\Ph_{n, \vec \xi}} f(\vec \xi) - \E_{\P_{\vec \xi}(\vec \theta)} f(\vec \xi)\right].
\end{equation}
The above display is reminiscent of the Wasserstein-GAN \eqref{eq:W-GAN} \cite{liu2023robust}, where $d_{\text{W}}$ and its variational representation are employed instead of $d_{\text{TV}}$. In $\phi$-GANs, TV-GANs, and W-GANs, the involved function spaces, such as $\{f: \|f\|_{\infty} \le 1\}$ and $\{f: \|f\|_{\text{Lip}} \le 1\}$, are $\vec \omega$-parameterized by proper neural networks using well-designed activation functions. Hence, computationally, maximizing over the discrimination functions $f_{\vec \omega}(\vec \xi)$ can be realized by maximizing over the parameters $\vec \omega$, leading \eqref{eq:W-GAN} and \eqref{eq:TV-GAN-practice} to a unified form of GAN as follows:
\begin{equation}\label{eq:TV-GAN-practice-compute}
\min_{\vec \theta \in \R^m} \max_{\vec \omega} \left[\E_{\Ph_{n, \vec \xi}} f_{\vec \omega}(\vec \xi) - \E_{\P_{\vec \xi}(\vec \theta)} f_{\vec \omega}(\vec \xi)\right].
\end{equation}
GANs have shown their excellent power in data generation and augmentation \cite{goodfellow2014generative}, as well as robust parameter estimation \cite{gao2019robust,liu2023robust}; for a specific WSC application, see \cite{balevi2020high}.

\subsection{Costs of Robustness}
Robust design inevitably introduces several costs that occur during real-world deployment. First, robustness often entails a trade-off against performance under nominal conditions. When uncertainties do not materialize, which is common in practice since many disturbances are random, robust solutions may sacrifice optimality relative to designs specifically tuned for the nominal conditions. Second, robust techniques typically incur additional computational burdens, as they must account for perturbations and seek worst-case optimality. Third, robust design serves as a remedial solution that tolerates uncertainties when new domain knowledge or additional data are unavailable. If such information can be acquired, adaptive methods can generally outperform robust methods, because the extra information reduces uncertainty. Therefore, the price of robustness is not only performance degradation under nominal conditions and increased computational cost, but also the opportunity cost of foregoing adaptive methods that can perform better when more information is available.

\section{Applications of Robust Methods in WSC}
In this section, recent advances in robust signal processing techniques for WSC are specifically exemplified. The aim is to show how general robust statistical, optimization, and machine learning approaches can be adapted to solve WSC problems.

\subsection{Robust Ranging-Based Localization} 
This subsection reviews an outlier-robust target localization scheme using range measurements \cite{wang2021denoising}; see also a similar investigation in \cite[p.~67]{zoubir2012robust}. For a $2$-dimensional planar localization problem, letting $(x, y)$ denote the position of the target and $(x_i, y_i)$ the position of the $i^\th$ sensor, the range $\rscl r_i$ from the target to the $i^\th$ sensor is
\[
\rscl r_i = \sqrt{(x - x_i)^2+(y - y_i)^2} + \rscl v_i,
\]
where $\rscl v_i$ is the zero-mean ranging noise of the $i^\th$ sensor. An outlier-robust M-estimation scheme for target localization, under a well-designed cost function $\rho$, is
\[
\min_{(x, y)} \sum^n_{i = 1} \rho\left(\rscl r_i - \sqrt{(x - x_i)^2+(y - y_i)^2}\right);
\]
recall \eqref{eq:M-estimation}, \eqref{eq:estimation-mean-cov}, and \eqref{eq:M-estimation-mean-cov}. The influence function $\d\rho(t)/\d t$ plays a crucial role in outlier handling: that is, if it is monotonic, we damp large-valued outliers, while if it is redescending, we throw away outliers; see Fig. \ref{fig:re-descending} and \cite[Fig.~3]{wang2021denoising}. Essentially, every specific choice of $\rho(\cdot)$ corresponds to an implicit assumption of the distribution of $\rscl v_i$ \cite{huber1964robsut}. In the usual case where $\rho(\cdot)$ is quadratic, it is assumed that $\rscl v_i$ is Gaussian and outlier-free. If instead, $\rho(\cdot)$ is the absolute value function, it is assumed that $\rscl v_i$ is Laplacian, i.e., fat-tailed and outlier-aware. Yet, if $\rho(\cdot)$ is the Huber's cost function in \eqref{eq:Lhuber-cost}, it is assumed that $\rscl v_i$ is Gaussian in the middle but has Laplacian tails \cite[p.~81]{huber1964robsut}.

\subsection{Adversarially Robust Sensing}
This subsection reviews the adversarially robust sensing techniques in \cite{modas2020toward}, which is a domain-specific implementation of the adversarial learning framework \eqref{eq:adversarial-learning} in modality sensing (i.e., pattern recognition via classification) for autonomous vehicles. Letting $f_{\vec x}(\vec \xi)$ be a modality sensing function (i.e., a classifier or detector), parameterized by $\vec x$, the adversarially robust sensing can be formulated as
\[
\begin{array}{cll}
  \displaystyle \min_{\vec x} & \displaystyle \frac{1}{n} \sum^n_{i=1} \max_{\vec \zeta_i \in \Xi} \bb I[f_{\vec x}(\vec \zeta_i) \ne f_0(\vec \xi_i)] \\
   \st  & c(\vec \zeta_i, \vec \xi_i) \le \epsilon, &\forall i,\\
\end{array}
\]
where $f_0(\vec \xi_i)$ returns the true class label of $\vec \xi_i$, and $\bb I[\cdot]$ returns $1$ if the argument is true and $0$ otherwise; $c$ denotes an uncertainty quantification method (e.g., a norm) and $\Xi$ the natural legitimate domain of samples. The samples $\{\xi_i\}_i$ in the context of autonomous vehicles can be ultrasound, radar, GPS, lidar, and camera signals. Other cost functions, such as cross-entropy, can serve as alternatives to $\bb I[\cdot]$.

\subsection{Robust Channel Estimation Using GANs}
This subsection reviews the robust channel estimation method in \cite{balevi2020high} using a Wasserstein GAN. In wireless communications, a base-band and narrow-band signal propagation law can be written as
\[
\rvec x = \mat H \rvec s + \rvec v,
\]
where $\rvec x$ denotes the received signal, $\mat H$ the channel matrix, $\rvec s$ the transmitted signal, and $\rvec v$ the channel noise that is usually assumed to be zero-mean Gaussian. Wireless communications aim to first estimate the unknown channel $\mat H$ using a set of pilot signals $\{(\vec x_i, \vec s_i)\}_i$ (i.e., training dataset), and then recover a new transmitted signal $\rvec s_{\text{new}}$ using the estimated channel $\math H$ and the received signal $\rvec x_{\text{new}}$ (i.e., the testing sample). This statistical inference procedure is modeled by \eqref{eq:receive-combining}, where $\mat W$ is called an estimator (or receiver) and is a function of $\math H$. To improve the robustness of channel estimation against possible uncertainties under $d_{\text{W}, p}$, either data scarcity or data perturbations, \cite[p.~21]{balevi2020high} employs a Wasserstein GAN to learn a channel generator $\mat H = \vec G(\vec z)$ that maps from lower-dimensional representations $\vec z$ to channels $\mat H$. Consequently, with the learned generator $\vec G$, the signal transmission law becomes $\rvec x = \vec G(\vec z) \rvec s + \rvec v$. Given pilot signals $\{(\vec x_i, \vec s_i)\}_i$, the optimal representation $\vec z^*$ is estimated via a $2$-norm-regularized least-squares approach \cite[p.~22]{balevi2020high}
\[
\vec z^* \defeq \argmin_{\vec z} \frac{1}{n} \sum^n_{i=1} \|\vec x_i - \vec G(\vec z) \vec s_i\|^2_2 + \lambda \|\vec z\|^2_2,
\]
where $\lambda$ is a regularization coefficient. As a result, the estimated channel is $\mat H^* = \vec G(\vec z^*)$. One technical benefit of this GAN-based channel estimation method is that $\vec z$ is much lower-dimensional than $\mat H$, thus more sample-efficient. Another benefit is the improved robustness against uncertainties; recall robust estimation using W-GANs \eqref{eq:W-GAN} and \eqref{eq:W-GAN-regression}.



\subsection{Distributionally Robust Receive Combining}
This subsection reviews distributionally robust receive combining techniques in \cite{wang2025distributionallycombining}, which aim to recover transmitted wireless-communication symbols. Under usual stationary linear Gaussian channels, a receive combining problem can be formulated as \eqref{eq:receive-combining}. Operationally, the unknown true distribution $\P_{0, \rvec x, \rvec s}$ is replaced with the pilot-based empirical distribution $\Ph_{n, \rvec x, \rvec s}$, leading to distributional uncertainty; recall the principle of robust estimation in \eqref{eq:rule-robust-estimation}. Hence, a distributionally robust receive combining scheme can be formulated as
\[
    \min_{\mat W} \max_{\P \in \cal B_{\epsilon}(\Ph_{n, \rvec x, \rvec s})} \Tr \E_{\P} [\rvec s - \vec W \rvec x][\rvec s - \vec W \rvec x]^\H,
\]
for some well-designed distributional sets $\cal B_{\epsilon}(\Ph_{n, \rvec x, \rvec s})$. To ensure computational efficiency, \cite{wang2025distributionallycombining} proposes a diagonal-loading uncertainty set, equivalently transforming the above display to a regularized empirical risk minimization \cite[Corollary~1]{wang2025distributionallycombining}
\[
    \min_{\mat W} \Tr \E_{\Ph_{n, \rvec x, \rvec x}} [\rvec s - \vec W \rvec x][\rvec s - \vec W \rvec x]^\H + \epsilon \Tr[\mat W \mat W^H],
\]
for which closed-form robust solutions exist. This manifests the power of DRO and regularization in robust machine learning. For complicated cases, such as nonlinear estimation in reproduced kernel Hilbert spaces and neural network function spaces, see \cite{wang2025distributionallycombining}. Therein, data augmentation techniques using noise injection to achieve robustness are also discussed; in addition, conditional equivalence among DRO, regularization, and data augmentation is remarked.

\subsection{Robust Waveform Design for Integrated Sensing and Communication}
This subsection reviews a robust waveform design scheme for integrated sensing and communication (ISAC) in \cite{wang2024robust}. In the conventional ISAC waveform design literature, it is assumed that the communication channels are exactly known. In \cite{wang2024robust}, this assumption is questioned, and a robust waveform design scheme against channel uncertainties is studied. Letting $\mat S$ denote the communication constellation points to be transmitted through an over-the-air MIMO channel, $\mat H$ the communication channel, $\mat X$ the MIMO transmit waveform, $\mat R$ a perfect-sensing beampattern matrix (i.e., best for sensing), and $L$ the number of snapshots, the sensing-centric robust ISAC waveform design problem can be formulated as
\[
\begin{array}{cl}
   \displaystyle \min_{\mat X} \max_{\mat H}  & \|\mat H \mat X - \mat S\|^2_F \\
   \st  & \mat X \mat X^\H / L = \mat R, \\
        &   \|\mat H - \math H\| \le \epsilon,
\end{array}
\]
where $\math H$ denotes an estimated (thus uncertain) channel matrix and $\|\cdot\|_F$ the matrix F-norm. In the above display, the objective defines the multi-user interference energy to be minimized to improve the communication quality, while the first constraint indicates that the waveform prioritizes ensuring perfect sensing performance. The resulting robust design is computationally difficult to solve due to the non-convexity in $\mat X$ and the high-dimensional nature of $\mat H$ in massive MIMO scenarios, and hence, ad hoc solution methods are developed \cite{wang2024robust}.

\subsection{Robust Federated Learning Under Adversarial Users and Noisy Communications}
This subsection reviews robust treatments against adversarial users and noisy communications in federated learning in \cite{pillutla2022robust} and \cite{ang2020robust}, respectively. Federated learning has been an active research area in the WSC community due to the need for collaborative training across edge devices (e.g., mobile communication devices) and distributed sensors. Typical federated learning aims to solve the following machine learning problem 
\[
\min_{\vec x} \sum^k_{i = 1} \alpha_i \E_{\Ph_{n_i, \vec \xi, i}} h(\vec x, \vec \xi),
\]
where $k$ denotes the number of clients, $\Ph_{n_i, \vec \xi, i}$ the local empirical distribution constructed using $n_i$ local training data of the $i^\th$ client, and $\alpha_i$ the influence (usually proportional to $n_i$) of the $i^\th$ client. At the fusion center, the arithmetic mean is usually used to aggregate the decisions from clients:
\[
    \vec x^* = \displaystyle \frac{\sum^k_{i=1} \alpha_i \vec x^*_i}{\sum^k_{i=1} \alpha_i},
\]
where $\vec x^*$ denotes the fused global decision and $\vec x^*_i$ the local decision of the $i^\th$ client.

However, the above aggregation rule is sensitive to corrupted or malicious updates from adversarial users because arithmetic means are non-robust to outliers. Hence, \cite{pillutla2022robust} uses the geometric median, which is outlier-robust \cite[p.~758]{diakonikolas2019robust}, for decision-fusion:
\[
\vec x^* = \displaystyle \argmin_{\vec x} \sum^k_{i=1} \alpha_i \|\vec x - \vec x^*_i\|_2,
\]
where $\|\cdot\|_2$ denotes vector $2$-norm. In this sense, any outlier-robust mean estimator \eqref{eq:weighted-robust-mean} can be an alternative to the geometric median; see, e.g., \cite{huber2009robust,diakonikolas2019robust,lai2016agnostic}; recall M-estimation techniques based on \eqref{eq:M-estimation-mean-cov}. Note that the geometric median also admits a form of the weighted mean of $\{\vec x^*_i\}_i$ as in \eqref{eq:weighted-robust-mean}, under the Weiszfeld algorithm. 

Another consideration in federated learning is that decisions may be contaminated during noisy communications between the center and the clients, in both downlink and uplink directions. Hence, a worst-case optimization (i.e., an adversarial training) scheme can be employed to combat these communication uncertainties \cite{ang2020robust}:
\[
\min_{\vec x} \sum^k_{i = 1} \alpha_i \max_{\vec \delta: \|\vec \delta\| \le \epsilon} \E_{\Ph_{n_i, \vec \xi, i}} h(\vec x + \vec \delta, \vec \xi),
\]
where $\vec \delta$ denotes the communication uncertainties, which are assumed to lie in a norm ball $\{\vec \delta: \|\vec \delta\| \le \epsilon\}$. That is, at the client $i$, the local problem to solve is 
\[
\min_{\vec x} \max_{\vec \delta: \|\vec \delta\| \le \epsilon} \E_{\Ph_{n_i, \vec \xi, i}} h(\vec x + \vec \delta, \vec \xi),
\]
which generates a robust local decision $\vec x^*_i$, to be fused at the center using the rule of arithmetic mean.

\section{Conclusions}
This tutorial-style overview article discusses the principles and methods of robustness. Specifically, first, the concepts related to robustness are formalized, for example, robust performance, robust solutions, distributional robustness, robustness measures, etc. Second, robust technical treatments in statistics, optimization, and machine learning are investigated. Third, the costs of employing robust design are highlighted (i.e., the sacrifice of nominal optimality, the computational burden, and the limited performance improvements compared to adaptive methods). Fourth, recent robust signal processing methods in WSC are reviewed, such as robust localization, robust channel estimation using general adversarial networks, robust receive combining, robust federated learning, etc.

\textit{Open Research Challenges}: The primary research challenges of the robustness theory and methods are two-fold:
\begin{itemize}
    \item The real-world operating performance of a robust method depends significantly on the chosen uncertainty set. For a given real-world problem, however, the uncertainty quantification (e.g., the uncertainty set and its radius $\epsilon$) for the involved uncertain factor $\vec \xi$ or $\P_{\vec \xi}$ needs to be specifically studied. On the one hand, the choice among moment-based sets, $d_{\phi}$-balls, $d_{\text{TV}}$-balls, $d_{\text{W}, p}$-balls, and others, relies on the ad-hoc characteristics of the problem of interest. There are no general golden rules to automatically determine the uncertainty quantification metric. On the other hand, the radius $\epsilon$ cannot be overly large or small. Otherwise, the resulting robust method would be too conservative or less effective, respectively.
    
    \item After robust modeling, i.e., the determination of the cost function $h(\vec x, \vec \xi)$ and the uncertainty quantification of $\vec \xi$ (NB: the property of $h$ also largely impacts the robustness), another challenge is to design a computationally efficient algorithm to solve the corresponding robust model. The complication arises if the dimensions of the decision $\vec x$ and the parameter $\vec \xi$ are high. Moreover, it is preferred if the algorithm has closed-form solutions because signal processing problems require real-time responses.
\end{itemize}

For every real-world application, the above challenges in robust modeling and computation demand dedicated efforts, paving the way for future research.

\bibliographystyle{IEEEtran}
\bibliography{References}









\end{document}